\documentclass[12pt]{iopart}
\pdfoutput=1

\usepackage{color}

\expandafter\let\csname equation*\endcsname\relax
\expandafter\let\csname endequation*\endcsname\relax

\usepackage{amsmath}
\usepackage{amssymb}
\usepackage{graphicx}
\usepackage{subfigure}
\usepackage{units}
\usepackage{cite}
\usepackage{hyperref}

\newcommand{\new}[1]{{#1}}

\begin{document}

\title{Non-normality, reactivity, and intrinsic stochasticity in
  neural dynamics: a non-equilibrium potential approach}
 
\author{Serena di Santo}
\address{Departamento de Electromagnetismo y F{\'\i}sica de la
  Materia e Instituto Carlos I de F{\'\i}sica Te\'orica y
  Computacional. Universidad de Granada.  E-18071, Granada, Spain}
\address{Dipartimento di Matematica, Fisica e Informatica, Universit\`a di Parma, 
  Parco Area delle Scienze, 7/A - 43124, Parma, Italy}

\author{Pablo Villegas} \address{Departamento de
  Electromagnetismo y F{\'\i}sica de la Materia e Instituto Carlos I
  de F{\'\i}sica Te\'orica y Computacional. Universidad de Granada.
  E-18071, Granada, Spain}

\author{Raffaella Burioni} \address{Dipartimento di Matematica, Fisica e Informatica, Universit\`a di Parma, 
  Parco Area delle Scienze, 7/A - 43124, Parma, Italy} \address{INFN Sezione Milano Bicocca, Gruppo Collegato di Parma,
  Parco Area delle Scienze, 7/A - 43124, Parma, Italy}

\author{Miguel A. Mu\~noz} \address{Departamento de
  Electromagnetismo y F{\'\i}sica de la Materia e Instituto Carlos I
  de F{\'\i}sica Te\'orica y Computacional. Universidad de Granada.
  E-18071, Granada, Spain}

\begin{abstract}
  Intrinsic stochasticity can induce highly non-trivial effects on
  dynamical systems, \new{such as stochastic resonance,} noise induced
  bistability, and noise-induced oscillations, to name but a few. Here
  we revisit a mechanism --first investigated in the context of
  neuroscience-- by which relatively small intrinsic (demographic)
  fluctuations can lead to the emergence of avalanching behavior in
  systems that are deterministically characterized by a single stable
  fixed point (up state). The anomalously large response of such
  systems to stochasticity stems \new{from} (or is strongly associated
  with) the existence of a ``non-normal'' stability matrix at the
  deterministic fixed point, which may induce the system to be
  ``reactive''.  By employing a number of analytical and computational
  approaches, we further investigate this mechanism and explore the
  interplay between non-normality and intrinsic stochasticity. In
  particular, we conclude that the resulting dynamics of this type of
  systems cannot be simply derived from a scalar potential but,
  additionally, one needs to consider a curl flux which describes the
  essential non-equilibrium nature of this type of noisy non-normal
  systems. \new{Moreover, we shed further light on the origin of the
  phenomenon, introduce the novel concept of ``non-linear
  reactivity'', and rationalize the observed values of avalanche
  exponents.}
 \end{abstract}

\maketitle
\tableofcontents{}
\newpage

\section{Introduction.}
Noise is well-known to have a number of remarkable effects in many
different physical, chemical, and biological systems, such that it can
dramatically alter the predictions of deterministic approaches;
\new{noise-induced transitions, stochastic resonance, coherence
  resonance, and stochastic amplification of fluctuations are just a
  few well-acknowledged examples}
\cite{induced,Ojalvo-review,Sagues,McKane,Wallace,Hidalgo,Biancalani,Genovese}. Here,
we are ultimately interested in highly non-trivial stochastic effects
emerging in concomitance with non-standard deterministic dynamics,
i.e. with non-normal (linearized) dynamics as described below. For the
sake of specificity and clarity, we focus on a problem in the field of
neural dynamics, but the ideas and mechanisms discussed here go beyond
such an example and might be relevant in other apparently unrelated
contexts, such as e.g. theoretical ecology and population dynamics,
but, \new{for the sake of concision}, let us discuss in detail only
the case of neural systems.

The human brain exhibits persistent intrinsic activity even in the
absence of any stimulus or task \cite{restless}. Understanding the
origin and nature of such an irregular dynamics is a key challenge in
neuroscience \cite{Fox,Arieli,Deco,Schuster}. Pioneering experiments
by Beggs and Plenz, revealed the existence of scale-invariant episodes
of spontaneous electrochemical activity in neural tissues \emph{in
  vitro}, thereafter named \emph{neural avalanches}. Subsequently,
neural avalanches were detected in a wide range of experimental
settings, tissues and species both \textit{in vitro}
\cite{BP2003,Plenz2007,Beggs2008,Beggs2012,Arcangelis2012} and
\textit{in vivo}
\cite{Petermann2009,Hahn2010,Palva2012,Plenz2015-pyramidal}.
Experimental results produce rather consistently avalanche exponents
compatible with those of an unbiased branching process
\cite{BP2003,MAM-review}. Such a scale-invariant organization has been
taken as an indicator that cortical dynamics might operate close to a
critical state \cite{BP2003,Chialvo2010,Mora-Bialek,MAM-review}.  Such
a putative criticality might endow the system with huge sensitivity to
stimuli, large spatio-temporal correlations, optimal transmission of
information, and a number of other functional advantages
\cite{Schuster,MAM-review}. Theoretical models have actually proposed
a link between cortical dynamics and dynamical criticality
\cite{Millman2010,Levina2009,Levina2007,Bonachela2010,Rubinov2011,LG-PNAS}.

Nevertheless, the so-called criticality hypothesis in cortical
networks is still controversial \cite{Touboul2010,Touboul2} and some
authors have highlighted that it is not clear whether the available
empirical evidence actually calls for criticality or other alternative
explanations --such as noise, bistability, neutral dynamics, etc--
could be invoked \cite{Beggs2012,SOB,neutral-neural,Touboul2} (we
refer to \cite{MAM-review} for an extended overview and discussion of
these issues).  Thus,a careful scrutiny of possible alternative
scenarios for the emergence of scale-invariant avalanches is an
important general task. One of the main goals of the present paper is
to contribute to this broad open problem.

In mathematical and computational models, the emergence of irregular
bursts of neural activity in cortical tissues has often been related
to the large \emph{excitability} of the underlying dynamical system,
\new{as a result of which} the intrinsic dynamics can result in large
excursions from stable equilibria \cite{Ojalvo-review,Izhikevich}. In
particular, in models of neural networks, large excitability is often
been explained as resulting from a delicate interplay between
excitation and inhibition \cite{vVS,Brunel2000,Lim} that imposes that
parameters in mathematical models need to be fine tuned to set the
system close to the transition (critical) point between two distinct
phases (excitation-dominated and inhibition-dominated, respectively).

Within this context, Benayoun et al. made the very interesting
observation that avalanches, i.e. scale-invariant bursts of activity,
may emerge in finite populations of neurons in a rather robust way,
even when parameters are chosen away from the perfectly balanced
(critical) condition \cite{Benayoun}.  To understand this --and
inspired by previous work by Murphy and Miller \cite{MurphyMiller}--
they relied on a mechanism called ``balanced amplification of
fluctuations'' \cite{Benayoun}, which is relevant for noisy neural
systems when the levels of excitation and inhibition are relatively
balanced, i.e. they are similar but not necessarily identical (in
other words, when the system does not necessarily lie at the very
critical point (see below)).

Our first aim here is to shed further light on the various factors
contributing to the appearance of non-critical --but still
approximately scale-invariant-- avalanches of activity emerging under
balanced amplification of fluctuations \cite{Benayoun}.

The structure of the paper is as follows. In Section 2 we briefly
review the basic aspects of the dynamical model for neural systems,
and describe in some detail the mechanism uncovered by Benayoun et
al. for the emergence of avalanching behavior in finite
populations. In Section 3 we revisit the results in \cite{Benayoun},
putting the emphasis of the importance of the interplay between
``non-normal forms'',``reactive dynamics'' and ``demographic
fluctuations'' in neuronal populations. Moreover, we construct a
statistical-mechanics description of the problem based on
non-equilibrium potentials which helps shedding light onto the overall
phenomenology. Finally, we present a broader discussion of our
results.

\begin{center}
\begin{figure}[h]
\centering{} \includegraphics[width=0.80\columnwidth]{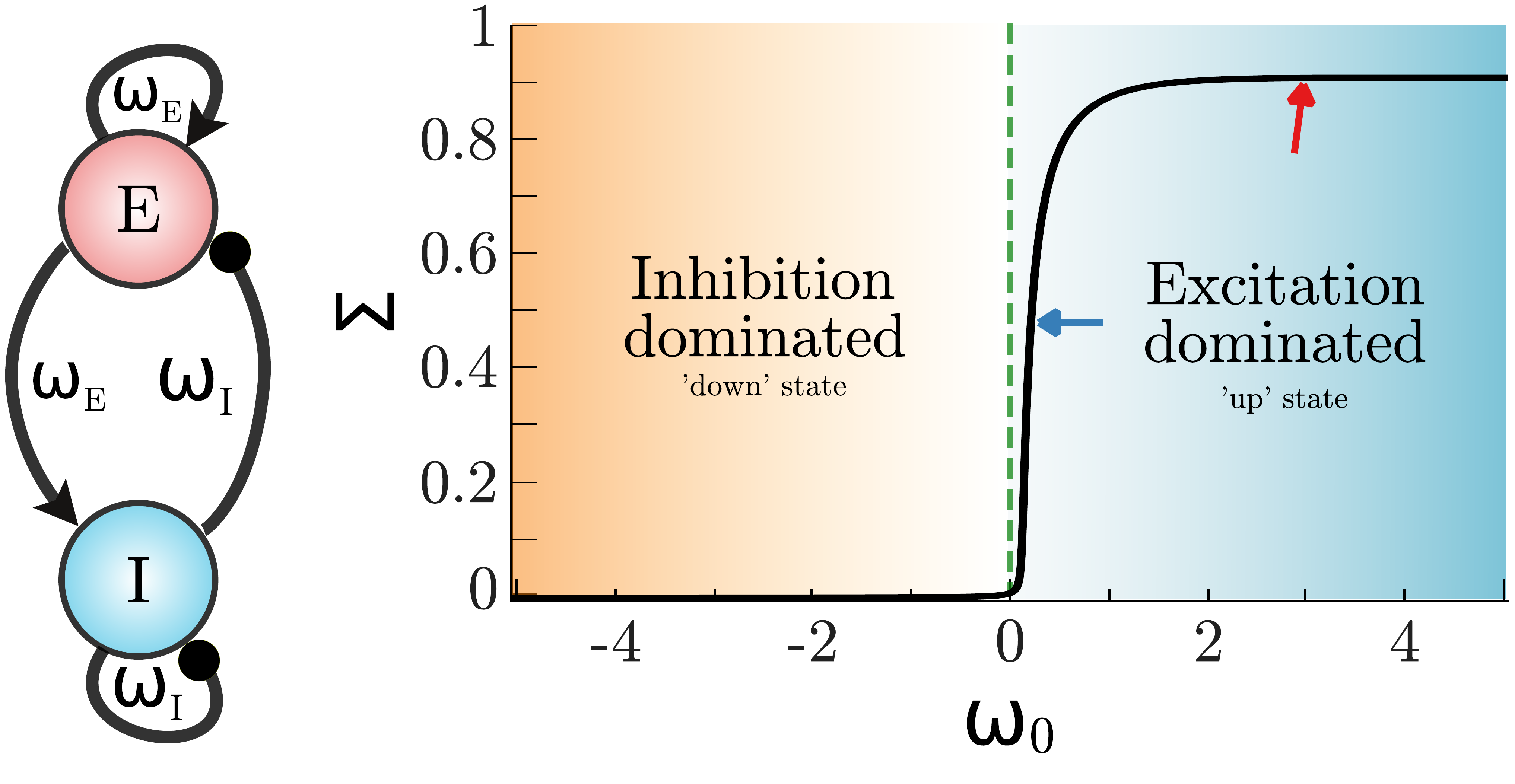}
\caption[Sketch of the Wilson-Cowan model]{Left figure: sketch of the
  Wilson Cowan model including an excitatory and an inhibitory
  population.  The excitatory population interacts with coupling
  parameter, $\omega_{E}$ and the inhibitory population with coupling
  $\omega_{I}$ (a more general case is that in which the effect of
  excitation on excitatory and inhibitory populations, is different
  from the effect of inhibition).  Right: Phase diagram; the phase of
  the system depends on the control parameter, ie. with the difference
  $\omega_{0}=\omega_{E}-\omega_{I}$. If $\omega_{0}>0$ the system is
  in the active or ``up'' phase, while for negative values it is in
  the low-activity ``down'' state. The blue and red arrows signal two
  specific choices of values that will be analysed in detail in what
  follows.}
\label{fig:phasetr}
\end{figure}
\par\end{center}

\section{Wilson-Cowan model for excitatory and inhibitory networks}

Following Benayoun et al. \cite{Benayoun}, we consider the
Wilson-Cowan mean-field description of a large-scale homogeneous
population of excitatory and inhibitory neurons \cite{Wilson1972}. As
usual, in this type of mean-field approaches, the connections between
neurons within a large population are assumed to be dense enough so
that heterogeneity and fluctuations can be neglected.  In spite of its
simplicity, the Wilson-Cowan model encompasses --as parameter values
are changed-- a plethora of possible scenarios, which strikingly
resemble experimentally observed dynamical regimes of neural dynamics,
such as multiple coexisting stable states (Up-Down states),
oscillatory behavior, simple and multiple hysteresis loops, etc.
\cite{Wilson1972,BorisyukKirillov,HoppensteadtIzhikevich,36WC}.  We
refer to Eef. \cite{WC-review} for a detailed analysis of the Wilson
Cowan model and its rich phenomenology.

For the particular version of the Wilson-Cowan model considered in ref.
\cite{Benayoun} (sketched in Fig. \ref{fig:phasetr}A), the mean-field equations
describing the overall dynamics of the activity (density of active
neurons) for the two subpopulations of excitatory ($E$) and
inhibitory ($I$) neurons, respectively,  read \cite{Wilson1972}:
\begin{equation}
\begin{cases}
\displaystyle{\frac{dE}{dt}}=-\alpha
E+\left(1-E\right)f\left(s\right)\\
\nonumber \\
\displaystyle{\frac{dI}{dt}}=-\alpha I+\left(1-I\right)f\left(s\right),
\end{cases}\label{eq:WC_det}
\end{equation}
where $\alpha$ is the rate of spontaneous activity decay, 
$s$ is the averaged incoming current 
\begin{equation}
\ensuremath{s=\omega_{E}E-\omega_{I}I+h},
\end{equation}
which is simply the sum of all synaptic inputs, both excitatory and
inhibitory,  weighted by their 
synaptic efficacies ($\omega_{E}$ and $\omega_{I}$, respectively), 
plus an external small constant input current $h$, and $f(s)$ is a
sigmoid response function:
\begin{equation}
f\left(s\right)=\begin{cases}
\tanh\left(s\right) & s\geq0\\
0 & s<0.
\end{cases}
\end{equation}
This set of equations exhibits a regime shift at
$\omega_{0} \equiv \omega_{E}-\omega_{I}=0$, from an inactive (inhibition
dominated) to an active (excitation dominated) phase. This is
illustrated in Fig.\ref{fig:phasetr} where we plot the averaged global
stationary activity, $\Sigma=(E+I)/2$ as a function of $\omega_{0}$,
while keeping $\omega_{s} \equiv \omega_{E}+\omega_{I}$ constant. In
particular, fixing $h=0$ the regime shift is a true bifurcation
occurring at the ``critical'' transition point
$\omega_0=0$\footnote{While, if $h\neq0$,  there is a rapid regime
  shift, but without a true discontinuity in the derivative at the
  transition point, i.e. without a true phase transition. }.

To go beyond this simple mean-field picture --describing
infinitely-large neural populations-- Benayoun et al. considered a
large but finite population of binary neurons obeying some standard
dynamical rules \footnote{In particular, they consider a
  fully-connected network model of individual spiking neurons such
  that (i) each neuron is either active or quiescent, (ii) the
  probability that each quiescent neuron becomes active depends on (a
  sigmoid function of) its total synaptic input and (iii) each active
  neuron relaxes to the quiescent state at a constant rate
  \cite{Benayoun} (see also \cite{Fanelli} for a similar
  approach).}. Starting from such microscopic (rate) model and
employing a large-system expansion \cite{Gardiner,vanKampen}, Benayoun
et al. were able to recover the above deterministic Wilson-Cowan
dynamics, Eq. (\ref{eq:WC_det}), in the infinite-size limit. But,
additionally, considering next-to-leading-order corrections, they were
also able to explicitly determine the stochastic term to be added to
Eq.  (\ref{eq:WC_det}) accounting for the finite size of the
population, leading to the set of stochastic equations
\begin{equation}
\begin{cases}
  \displaystyle{\frac{dE}{dt}}=-\alpha E+\left(1-E\right)f\left(s\right)+
\sqrt{\alpha E+\left(1-E\right)f\left(s\right)}\eta_{E}\\ \nonumber
\\
  \displaystyle{\frac{dI}{dt}}=-\alpha
  I+\left(1-I\right)f\left(s\right)+\sqrt{\alpha
    I+\left(1-I\right)f\left(s\right)}\eta_{I}
\end{cases}\label{eq:WCNoise}
\end{equation}
where $\eta_{E,I}$ are uncorrelated Gaussian white noises, with
network-size dependent amplitude $\sigma \propto 1/\sqrt{N}$,
interpreted in the Ito sense \cite{vanKampen}.

\begin{figure}
\centering{} \includegraphics[width=0.98\columnwidth]{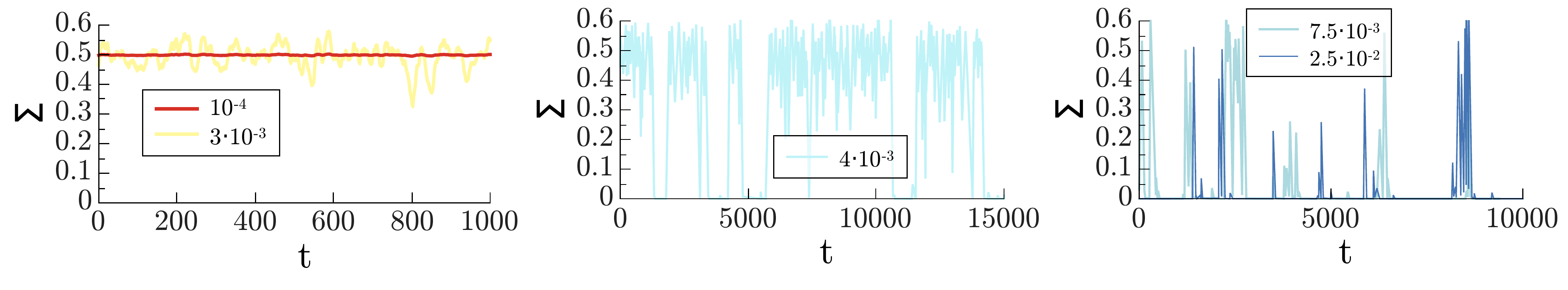}
\caption[Temporal series of the system activity showing avalanches under a ``balance
amplification'' condition]{Timeseries of the total excitatory and
  inhibitory  activity $\varSigma=(E+I)/2$ in a balanced
  condition ($\omega_{E}=7$, $\omega_{I}=\frac{34}{5}$; blue arrow in
  Fig.1) for increasing levels of noise (from left to right, as
  marked in the figure legends). Above a certain noise-amplitude threshold the
  system reaches the inhibition-dominated state and exhibits
  avalanching behavior. Parameter values:  $\alpha=0.1, h=10^{-3}$
  \label{fig:timeseriesWC}}
\end{figure}

\subsection{Model phenomenology: computational results}

By performing computational simulations of their
individual-neuron-based model, Benayoun et al. \cite{Benayoun}
observed that --when the difference between excitatory and inhibitory
synaptic weights, $\omega_0$, is small with respect to their sum
$\omega_{0}\ll\omega_{s}=\omega_E+\omega_I$; i.e. in a situation that
is termed ``balanced'' \cite{Benayoun,MurphyMiller}, the actual
dynamics departs quite dramatically from the mean-field expectations,
even for relatively low noise amplitudes (i.e. for large but not
infinite network sizes).  \new{In particular, and quite remarkably,}
for parameter values ($\omega_0 \gg 0$) such that the mean-field
Wilson-Cowan equations predict a stable up state, the dynamics --in
the presence of noise-- turns out to spend most of the time close to
the down state (with very low activity) \new{while showing} frequent
bursts of activity (such as those illustrated in Fig.2).

The sizes $S$ and durations $T$ of such bursts or avalanches of
activity turn out to be distributed as (approximated) power-laws,
\new{similar to those observed experimentally in actual neuronal
  networks}, even if the exponent values are found to be
detail-dependent rather than universal; i.e. do not necessarily
coincide with the experimentally-observed branching process exponents
\cite{Benayoun}. Thus, the problem of determining the precise origin
and values of these exponents remains open.

\begin{figure}
  \centering{} \includegraphics[width=0.98\columnwidth]{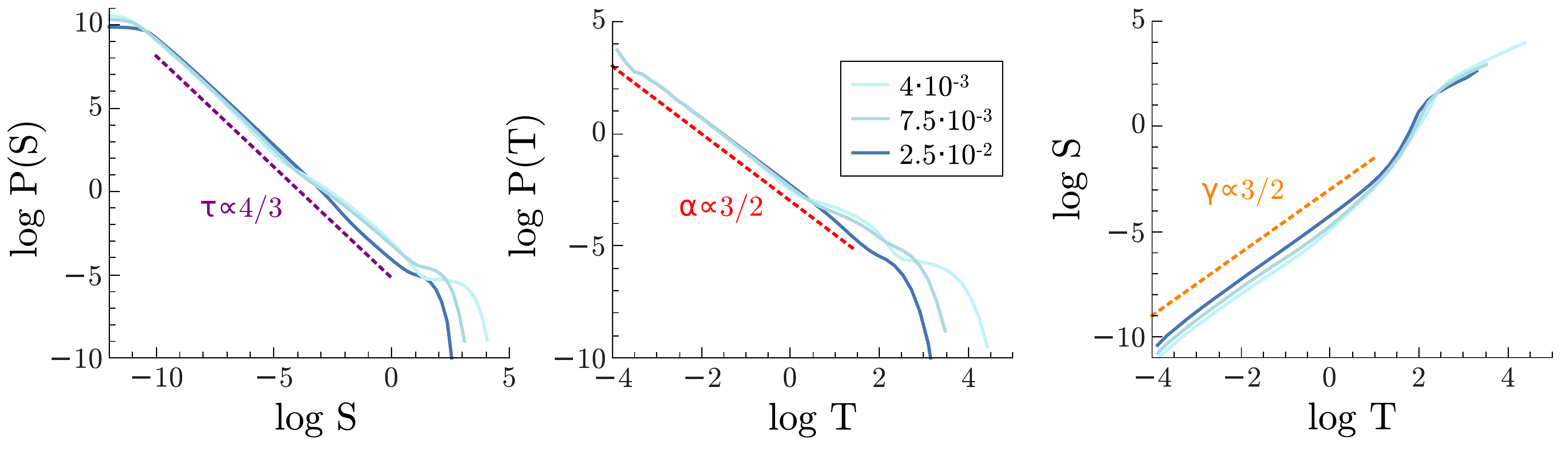}
  \caption{(Left) Avalanche size distribution and (Center) avalanche
    time distribution for different levels of noise ($\sigma$, see
    Legend in the central plot).  Reasonably good fits to power-laws
    are obtained employing the well-known values for the return times
    of a random walk (see e.g. \cite{logarithmicpot}); the quality of
    the fits improves with increasing noise, but in any case they are
    not perfect straight lines, revealing non-perfect scale
    invariance. (Right) The averaged avalanche size as a function of
    the duration $T$, also scales as predicted by the statistics of
    random walks, $S \sim T^{\gamma}$, with $\gamma=3/2$.}
\label{fig:powerlawsWC}
\end{figure}

To illustrate and \new{extend the above numerical analyses}, we
performed our own computer simulations but, opposite to Benayoun et
al., not of the individual-neuron-based model, but directly of the set
of stochastic equations, Eq.(\ref{eq:WCNoise}).  For the numerical
integration we used an Euler-Maruyama \new{scheme} with step size
$\Delta t=10^{-4}$ (and checked that results are not very sensitive to
this choice). In particular, we considered $\omega_E=7$ and
$\omega_I=34/5$, which is a balanced case in the sense that $\omega_0$
($1/5$) is relatively small when compared with $\omega_s$
($69/5$). Our results are summarized in Fig.\ref{fig:timeseriesWC}
where we confirm the existence of a stable fixed point with small
fluctuations around it for small noise amplitudes
(e.g. $\sigma=10^{-4}$).  On the other hand, slightly larger values of
noise amplitude (i.e. smaller system sizes) produce large fluctuations
around such a stable fixed point, and, more remarkably, above some
value of the noise amplitude, $\sigma \approx 5. 10^{-3}$, the
stability of the up state is severely compromised and the system ends
up hovering around a low-activity value with large excursions to the
up state. Such excursions closely resemble avalanches of activity,
growing, spreading and coming back to the inhibition-dominated down
state.

\new{The resulting avalanche-like dynamics suggests a definition of
  the duration $T$ and size $S$ of avalanches as the activity over a
  small arbitrary threshold, allowing us to confirm the results of
  ref. \cite{Benayoun}; i.e.  both avalanche duration and size follow
  an approximate power law of the form $P(T)\sim T^{-\alpha}$ and
  $P(S)\sim S^{-\tau}$ with $\langle S\rangle\sim T^{\gamma}$}
\footnote{These exponents obey the usual scaling relationship
  $\gamma=(\alpha-1)/(\tau-1)$ \cite{avalanches,logarithmicpot}.} (see
Fig.3).  Contrarily to the case of \cite{Benayoun} we can obtain
relatively clean values for the exponents. In particular, although the
size and time distributions are not perfect straight lines in the
double-logarithmic plot, revealing a lack of strict scale invariance,
the slopes of the best fits are compatible with the well-known
exponents for the return times of a random walk, i.e. $\alpha=3/2$,
$\tau=4/3$ and $\gamma=3/2$, rather than those of the branching
process ($\alpha=2$, $\tau=3/2$ and $\gamma=2$: see
e.g. \cite{logarithmicpot} \new{for a pedagogical discussion of these
  two classes}). \new{Let us finally remark that scale invariance is
  only approximate in all cases; there is no simple limit in which
  perfect power laws and scaling emerge.}

\begin{center}
\begin{figure}[h]
  \centering{} \includegraphics[width=0.88\columnwidth]{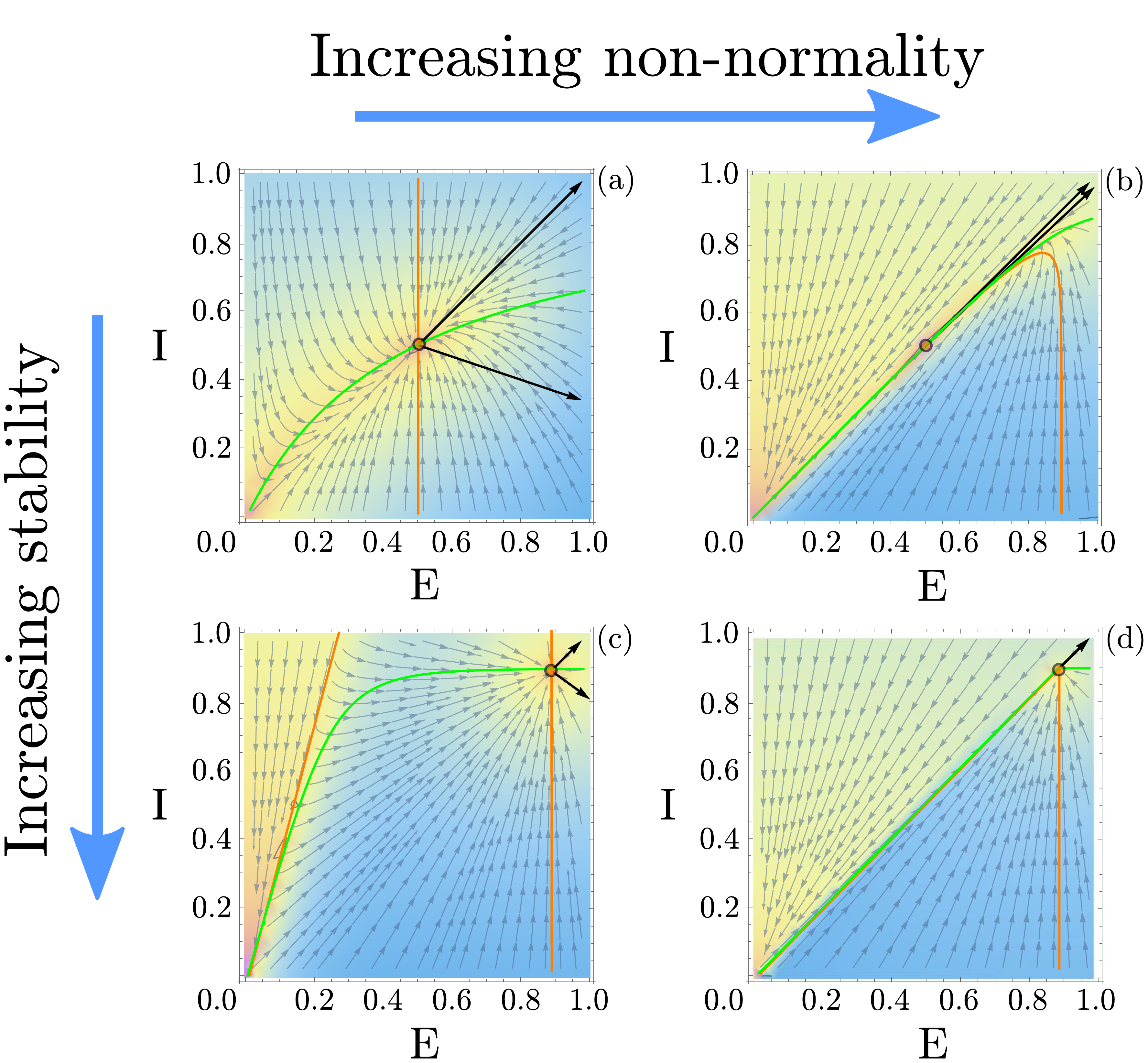}
  \caption[Phase portrait of the Wilson-Cowan model]{
    Excitation-inhibition (E-I) phase portraits for four different
    sets of parameter values; cases a and b have a relatively weak
    stability (eigenvalues of the Jacobian at the fixed point
    $E=I\simeq0.5$ are $\lambda_1=0.299, \lambda_2=0.202$) while cases
    c and d are strongly stable (eigenvalues $\lambda_1=-3.63$ and
    $\lambda_2=-3.13$ in the fixed point $E=I\simeq0.9$). On the other
    hand, there are cases with small (a and c) and large (b and d)
    non-normality and reactivity, respectively (see main text).  The only stable
    fixed point in each case --wherein the nullclines intersect-- is
    represented as a red circle and the corresponding eigenvectors are
    indicated by black arrows.  (a) Case with weak stability and
    relatively small non-normality $\mathcal{NN}=0.068$; parameters
    $\omega_{E}=\frac{1}{5}$, $\omega_{I}=0$. The system is not
    reactive, ${\cal{R}}=-0.18$ (see Appendix A). (b)
    Case of balanced-amplification
    (parameters $\omega_{E}=7$, $\omega_{I}=\frac{34}{5}$): the
    stability is very weak with respect to the non-normality $\mathcal{NN}=0.997$
    and reactivity is also relatively large ${\cal{R}}=3.13$. 
The nullclines are in close proximity to each other, in
    contrast with case a, and this induces a shear flow, with
    velocity fields pointing in opposite directions at
    either side of the diagonal.  (c) Case with relatively strong
    stability and small non-normality; $\mathcal{NN}=10^{-6}$ ($\omega_{E}=4$,
    $\omega_{I}=1$); the fixed point is non-reactive
    ${\cal{R}}=-1.093$.  (d) Case with strong linear stability at the
    fixed point and large non-normality, $\mathcal{NN}=0.706$ ($\omega_{E}=3000$,
    $\omega_{I}=2997$). The equilibrium is reactive ${\cal{R}}=0.106$.
The main observation to be made is that  the deterministic dynamics is
radically
different in the cases with and without a shear flow, i.e. with or
without a ``scar'' in the diagonal.
\label{fig:phaseport}}
\end{figure}
\par\end{center}
\subsection{Deterministic dynamics}

To shed light onto this remarkable noise-induced effect, one can
scrutinize the phase portrait of the deterministic dynamics of
Eq.(\ref{eq:WC_det}).  Figure \ref{fig:phaseport} shows the phase
portrait plane for four different sets of parameter values.  In the
upper panels (corresponding to parameters signaled with a blue arrow
in Fig.\ref{fig:phasetr}) the stability is weak ($\omega_0=0.2$) while
in the lower ones (red arrow in Fig.1) it is stronger
($\omega_0=3$). The most relevant case for our discussion here is that
in Fig.\ref{fig:phaseport}b, which corresponds to the balanced
condition.  One can readily observe that, in this last case, the force
(or velocity) field is rather anomalous: it exhibits a sort of
``scar'' close to the diagonal, with shear flows in opposite
directions above and below it, allowing for wild excursions far away
from the fixed point if it is perturbed \footnote{In other words, the
  vector field $(\vec{\dot{E}},\vec{\dot{I}})$ shows an abrupt jump
  around the diagonal, where the two nullclines are very close to each
  other, even if they intersect just at the only fixed
  point.}\cite{Benayoun}.  Similarly, in Fig.\ref{fig:phaseport}d,
even if the system is more stable, there also exists a scar allowing
for such type of trajectories.  On the contrary, in
Figs. \ref{fig:phaseport}a and \ref{fig:phaseport}c, the stability of
the system is quite strong and phase portraits are completely
different. We now discuss these features step by step, \new{starting
  from} a linear stability analysis.

\section{Results}
\subsection{Linearized deterministic dynamics: non-normality and reactivity}

To illustrate the phenomenon of non-normal dynamics (still following
\cite{Benayoun}), it is convenient to change variables to
$\Sigma=(E+I)/2$, $\Delta=(E-I)/2$, so that Eq.\ref{eq:WC_det} becomes
\begin{equation}
\begin{cases}
\displaystyle{\frac{d\varSigma}{dt}}= &
-\alpha\varSigma+\left(1-\varSigma\right)f\left(\theta\right)\\ \\
\displaystyle{\frac{d\varDelta}{dt}}= & -\varDelta\left(\alpha+f\left(\theta\right)\right)
\end{cases}\label{eq:WCSigma}
\end{equation}
with $\theta=\omega_{0}\Sigma+\omega_{s} \varDelta+h$. Clearly, these
equations can only have fixed points of the form $(\Sigma^*,0)$, i.e.
lying necessarily in the diagonal of the $(E,I)$ phase portrait.
\new{In particular, in all the considered cases (see Fig.4) there is a
  ``up-state'' fixed point with $E=I$ in which excited and inhibited
  neural popualtions coexist, and equilibrate each other.}

  A standard linear stability analysis around the (only) up-state
  fixed point leads to the Jacobian matrix
\begin{equation}
  J=\left(\begin{array}{cc}
            -\lambda_{1} & \omega_{ff}\\
            0 & -\lambda_{2}
\end{array}\right)\label{eq:Jacob}
\end{equation}
where the eigenvalues are
$\lambda_{1}=(\alpha+f(\theta^*))+(1-\Sigma^*)\omega_{0}f'(\theta^*)$
and $\lambda_{2}=(\alpha+f(\theta^*))$ and
$\omega_{ff}=(1-\Sigma^{*})(\omega_{E}+\omega_{I})f'(\theta^*)$,
\new{with $\theta^{*}=\omega_0\Sigma^{*}+h$} (where we have used
$\Delta^*=0$).  In what follows, the spontaneous decay $\alpha$ and
the spontaneous activation rate $h$ are fixed to relatively small
values (e.g.  $\alpha=0.1,\,h=10^{-6}$). In particular, this imposes
that if $\omega_0$ is small and positive, then, $\lambda_{1}$ and
$\lambda_{2}$ are small \cite{Benayoun}, as corresponds to a
\emph{weakly stable} fixed point. Therefore, as usual, the (weak)
stability of the fixed point is controlled by the \new{distance to}
the transition point (at $\omega_0=0$).

The very structure of the Jacobian makes it clear that diagonal terms
--the set of eigenvalues-- do not enclose all the information about
the linearized dynamics. The non-vanishing off-diagonal term (the
so-called \emph{feed-forward} term, $\omega_{ff}$, cannot be eliminated
by changing variables, and establishes a clear asymmetry in the
dynamics: $\Delta$ boosts $\Sigma$ but not the other way around).

Triangular matrices such as $J$ are a particular case of
\emph{non-normal} forms/matrices (also called non-self-adjoint
matrices), meaning that $J^{*}J\neq JJ^{*}$, where $J^{*}$ is the
conjugate transpose of $J$.  As a consequence, such matrices are not
diagonalizable through a unitary transformation, or in other words,
the associated basis of eigenvectors is not orthogonal
\cite{Horn,Trefethen}\footnote{Let us remark that the change of
  variables employed above, $(E,I)\rightarrow(\Sigma,\Delta)$, is what
  is usually called a ``Schur transformation'', generating a
  triangular matrix; indeed, the Schur decomposition is a
  decomposition of a given matrix $A$, such that $A=MTM^{-1}$ where
  $M$ is a unitary matrix and $T$ is an upper triangular matrix: the
  associated Schur form \cite{Horn}.}. Non-normal matrices might be
unfamiliar to many physicists, grown up with quantum mechanics, where
observables are represented by Hermitian (self-adjoint)
operators. However, non-normal forms have a long tradition in other
realms of physics such as turbulence, where they play an important
role in e.g. shear \new{and pipe flows}
\cite{Hydrodynamics,Magnetohydrodynamics,pipe,Trefe2}, in control
theory \cite{Uncertain}, in ecology \cite{Caswell}, lasers
\cite{Lasers}, and even, recently, in quantum mechanics
\cite{Q,non-HermitianQM}.
%%%%%%%%%%%%%%%%%%%%%%%%%%%%%%%%%%%%%%%%%%%%
 
In the case under scrutiny, the basis of eigenvectors in the variables
$(\Sigma,\Delta)$ is
\[
\left(\begin{array}{c}
1\\
0
\end{array}\right),\,\,\left(\begin{array}{c}
1\\
\xi
\end{array}\right)
\]
\new{with $\xi=\omega_{0}/(\omega_{E}+\omega_{I})=\omega_0(1-\Sigma^{*})f'(\theta^*)/\omega_{ff}$}, leading to almost
identical (degenerate) eigenvectors in the limit $\xi\rightarrow0$, \new{i.e. $\omega_{ff}\gg \lambda_{1,2}$, i.e. the balance condition}.
Also, importantly, the scalar product of both eigenvectors does not
vanish, i.e. they do not form an orthogonal basis (as illustrated in
Fig.\ref{fig:phaseport}, especially in cases b and d).
%, while cases a and c are close to normal.).

A large value of the feed-forward term $\omega_{ff}$ in
Eq.(\ref{eq:Jacob}) may induce a huge impact on the dynamics around a
weakly stable fixed point.  In particular, when the matrix $J$
operates on a small perturbation vector along the $\Delta$ direction,
$(0,\epsilon)$, applying the linearized dynamics once, gives
$(+ \omega_{ff} \epsilon, -\lambda_2 \epsilon)$ \new{which results in
  a (``contracting'') response in such a direction, but a much larger
  (``expanding'') outcome along the $\Sigma$ direction.}  This
illustrates that a large value of the feed-forward term $\omega_{ff}$
is able to strongly affect the linearized dynamics. Observe that this
can occur even when the eigenvalues are not close to zero (i.e. the
system is not necessarily close to the transition point) if the
feedforward term is large with respect to their \new{absolute value
  (see below)}.  As a matter of fact, since the eigenvalues depend on
the control parameters $\omega_E$ and $\omega_I$ \new{only through the
  combination $\omega_{0}=\omega_E-\omega_I$ --while $\omega_{ff}$
  only depends on $\omega_{s}=\omega_E+\omega_I$-- }the condition for
the appearance of the mechanism discussed above is the ``balance
condition'' $\omega_{0}\ll\omega_{s}$ \cite{MurphyMiller}.

It is noteworthy that it is possible to quantify the degree of
non-normality as the weight of the feed-forward interaction
(off-diagonal elements) with respect to that of the eigenvalues. In
particular, one can quantify the strength of the feedforward elements
of a given triangular matrix as:
\begin{equation}
\mathcal{NN}=1-\frac{{\sum_{i}|\lambda_{i}|^{2}}}{\sum_{n}m_{n}^{2}}
\end{equation}
\new{where $\lambda_{i}$ are the diagonal elements (eigenvalues), and
  the term in the denominator is the sum of the squares of all matrix
  elements, $m_n$. In our particular case, this implies that, fixing
  $\omega_{0}$, $\mathcal{NN}$ tends to grow with $\omega_{s}$}. In
other words, the balance condition $\omega_{0}\ll\omega_{s}$ implies
that the strength of non-normality is large. In particular, in Fig.4
cases b and d are highly non normal (non-normality
$\mathcal{NN}=0.997$ and $\mathcal{NN}=0.706$, respectively), while a
and c are hardly non-normal ($\mathcal{NN}=0.068$ and
$\mathcal{NN}=10^{-6}$, respectively).

The anomalous behavior we have described in association with
non-normal matrices is sometimes called ``reactivity'' in the
(theoretical ecology) literature \cite{Caswell}. \new{As carefully
  explained in Appendix A,} reactivity describes the property of
linear (or linearized) stable systems such that their dynamics --even
if converging asymptotically to a stable fixed point-- can exhibit
unusually long-lasting transient behavior. In other words, the system
can be strongly driven away from the fixed point (actually increasing
the modulus of the perturbation vector) before converging to its
steady state. Also, it is shown in Appendix A that all reactive
matrices are non-normal, but the opposite is not true. In particular,
in the examples of Fig.4 all stability matrices are non-normal, but
only (b) and (d) are also reactive.

%%%%%%%%%%%%%%%%%%%%%%%%%%%%%%%%%%%%%%%%%%%%%%

\subsection{Linearized dynamics with noise}
Thus far we have revisited results in \cite{MurphyMiller} and
\cite{Benayoun} regarding the very peculiar features of the
deterministic dynamics in systems with non-normal stability matrices.
In order to proceed further in our analytical understanding of the
avalanching phenomenon, we now study the stochastic dynamics (up to
leading-order approximation) around the deterministic linearized
equations, to explore the role of the so-introduced fluctuations
around the deterministic fixed point.  For this, we consider the set
of linearized Langevin equations
\begin{equation}
\left(\begin{array}{c}
\dot q_1\\
\dot q_2
\end{array}\right) = 
\left(\begin{array}{cc}
-\lambda_1 & \omega_{ff}\\
0 & -\lambda_2
\end{array}\right)
\left(\begin{array}{c}
q_1\\
q_2
\end{array}\right)
+\sigma
\left(\begin{array}{cc}
d_1 & 0\\
0 & d_2
\end{array}\right)
\left(\begin{array}{c}
\eta_1\\
\eta_2
\end{array}\right)
\\
\label{eq:LinearLang}
\end{equation}
where the vector state $q=(q_1,q_2)$ describes deviations with respect
to the deterministic fixed point \new{$(\Sigma^*,\Delta^*=0$),}
i.e. $q_1=\Sigma-\Sigma^*$ and $q_2=\Delta$, $d_1$ and
$d_2$ are non-trivial functions depending on the parameters and on
\new{$(\Sigma^*,0$).}  For the sake of generality, we prefer to use
the following compact notation
\begin{equation}
 \dot{q}^\nu=F^\nu + \sigma G^{\mu\nu} \eta_\mu,
\end{equation}
where the vector force $F$, i.e. the vector with components
$F^{\mu}$ with $\nu=1, 2$, \new{is} $F = J q$, \new{$~~G$} is the
(diagonal) matrix (with constant noise amplitude evaluated at the
fixed point) and $\eta$ is a vector composed of two uncorrelated white
Gaussian variables (as usual, repeated indices are implicitly summed
over).

The above Langevin equation is equivalent to a
Fokker-Plank equation\cite{Gardiner}:
\begin{equation}
 \partial_t P = [-\partial_\mu F^\mu +
 \frac{\sigma^2}{2} \partial_{\nu\mu}B^{\nu\mu}]P
\label{FP}
\end{equation}
where \new{$B^{\mu\nu}=G^{\mu}_\gamma G^{\gamma}_\nu$}. Eq.\ref{FP} as usual, can be written
as a continuity equation:
\begin{equation}
 \partial_t P = -\partial_\mu j^\mu
\end{equation}
where the components of the current $j$ are
\begin{equation}
  j^\nu=F^\nu P - \frac{\sigma^2}{2} (B^{\mu\nu}\partial_\mu P - P \partial_\mu B^{\nu\mu}).
\label{eq:J}
\end{equation}

An overall stationary solution exists whenever the current is divergence
free; $\partial_\mu j^\mu=0$.  In general, if in the steady state the
current is null, $j^\mu=0$, detailed balance holds, and the system is
symmetrical under time reversal transformations: the stationary state
is at equilibrium. Otherwise, even in stationary conditions, there is
a net probability current flowing through the system, the detailed
balance is explicitly violated and the system is away from
equilibrium.

\new{Observe that the above Fokker-Planck equation corresponds to a
  multivariate Ornstein-Uhlenbeck process and, as such, can be solved
  exactly \cite{Gardiner,Risken} even in the case of non-vanishing
  stationary probability current \cite{Dotsenko,Wu,Puglisi}.  However,
  in order to illustrate the general procedure that shall be used
  afterwards (for more complex, non-linearized cases), in what follows
  we present an analytical approach based on path integrals and a
  weak-noise approximation to solve it.}

\new{Employing a simple (customarily employed) ansatz for the
  stationary solution in the weak-noise limit
  \cite{Graham1,Graham,Wio,Ja1,Ja2}, one can write}
\begin{equation}
 P_{st} \equiv  C \exp[-V(q)/\sigma+O(\sigma)],
\label{ansatz}
\end{equation}
where $C$  is a normalization constant and $V(q)$ a scalar potential,
one can  rewrite Eq.(\ref{eq:J}) --up to leading order in $\sigma$--  as:
\begin{equation}
 F^\nu=-\frac{1}{2} B^{\mu\nu} \partial_\mu V + j^\nu/P_{st}
 \label{eq:F}
\end{equation}
showing that the actual force in the Langevin dynamics can be
decomposed as the gradient of a scalar potential, plus an additional
force.  Note that such an additional term, $j^\nu/P_{st}$, can be seen
as a curl flux force, as it is necessarily divergence free in the
steady state ($\partial_\mu j^\mu=0$); thus, it has neither sources
nor sinks \footnote{Observe that this decomposition into a component
  which derives from a scalar potential plus a curl flux vector (which
  eventually can be written as the rotor of a vector potential) is
  nothing but the usual Helmholtz decomposition of the deterministic
  force into a curl-free and divergence-free components
  \cite{Arfken}.  }.
\begin{figure}[h]
\begin{center}
\vspace{-0cm}
   \includegraphics[width=0.7\columnwidth]{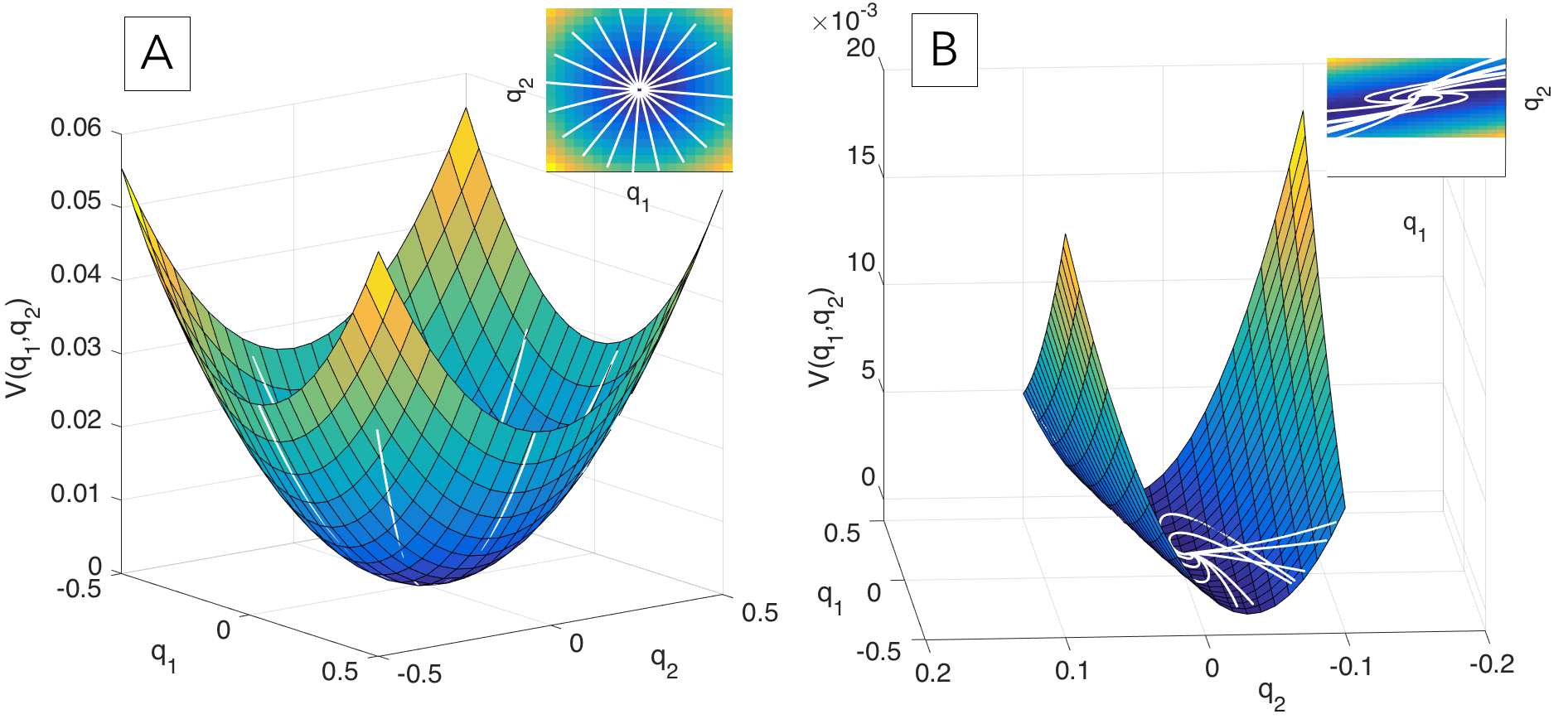}
   \caption{Non equilibrium potentials for the linearized dynamics, as
     calculated following \cite{Graham}. White curves are some of the
     points calculated using $V(q)=\int_{0}^{t^*} p \dot q dt$ along
     classical (or Hamiltonian) paths starting at the deterministic fixed-point, the
     surface represented in color code is a fit. The inset is the same
     plot, projected in the plane $V=0$. Left panel: for the case
     $\omega_{ff}=0$, the linearized dynamics is normal and the potential can be
     calculated analytically $V=\lambda_1x^2/2+\lambda_2y^2/2$ and
     $F=-\sigma B\nabla V/2$ (the force is conservative; i.e. it
     derives from the scalar potential $V$). Right panel $w=10$, the
     potential becomes much narrower in the y-direction.
     $\lambda_1=0.1, \lambda_2=0.11$. The projection in the inset
     shows that most-probable paths are much more curved and
     anisotropic than in the other case. In this case, the system
     dynamics do not simply follows the gradient of the scalar
     potential, but it has a curl component.
     \label{fig:pot}}
\end{center}
\end{figure}

In order to determine explicitly the two components of the potential
for Eq.(\ref{eq:LinearLang}), we first evaluate the scalar potential,
$V$, by making use of standard methods of the non-equilibrium
potentials literature relying on path integrals in the weak-noise
limit \cite{Graham,Graham1,Wio,Ja1,Ja2} and subsequently we evaluate
the curl flux force through Eq.(\ref{eq:F}).

In particular, plugging Eq.\ref{ansatz} into the Fokker-Planck equation,
the problem of finding the stationary solution up to leading order in
$\sigma$ implies solving
\begin{equation}
 \mathcal{H}(q,p)=F^\nu(q) p_\nu + \frac{1}{2} B^{\mu\nu} p_\mu p_\nu =0,
\end{equation}
where 
\begin{equation}
p_\mu \equiv\partial_\mu V
\label{p}
\end{equation}
which is formally identical to a Hamilton-Jacobi equation in classical
mechanics, being $\mathcal{H}(q,p)$ a Hamiltonian.

From a path integral viewpoint (see e.g. \cite{Graham,Wio}), the
probability to reach a given point from the deterministic attractor
can be expressed as $P(q) \propto \exp(-S_{min}(q))$, where
$S_{min}(q)$ is the minimum action along a ``classical'' trajectory
following the Hamilton's equations of motion,
$\dot q^\mu=\partial\mathcal{H}/\partial p_\mu$, such that starting at
the deterministic fixed point $q_{\min}$ (and with a value of $p$ as
close as possible to $0$ \footnote{Taking strictly $p=0$ trajectories
  do not leave the deterministic attractor, thus, vectors $p$ with
  very small but not vanishing modulus --and varying directions-- need
  to be considered.}) reaches an arbitrary point $q$ at time
$t^*$. Thus, the associated potential can be expressed (integrating
Eq.\ref{p} along such a path) as
\begin{equation}
V(q)=S_{\min}=\int_{q_{\min}}^q p ~dq=\int_{0}^{t^*} p ~\dot q~ dt
\end{equation}
and one just needs to compute this integral.

In Fig.\ref{fig:pot} we plot the potential that we have obtained by
numerically computing the above action for many classical paths
starting at the deterministic fixed point with different values of
  $p$  (with small modulus) pointing in different directions. \new{Such
  different paths reach different points $q$ \footnote{A linear
    interpolation algorithm has been employed to obtain a continuous
    curve, for visualization purposes} spanning the whole space of $q$
  values, allowing us to reconstruct $V(q)$
  \cite{Graham}.}

On the left of Fig.\ref{fig:pot} we represent the case with
$\omega_{ff}=0$, where the force is conservative, the resulting
potential has a standard paraboloid shape centered at the origin.  On
the other hand, in the case $\omega_{ff} \neq 0$ (in particular,
$\omega_{ff}=10$), the classical trajectories are much more curved,
and the resulting potential becomes asymmetric, and much
narrower along the direction transversal to the diagonal, as
illustrated in the right panel of Fig. \ref{fig:pot}. This reveals
that fluctuations along the diagonal are strongly more likely than
perpendicular ones.

\begin{figure}[h]
\begin{center}
\vspace{-0cm}
   \includegraphics[width=0.7\columnwidth]{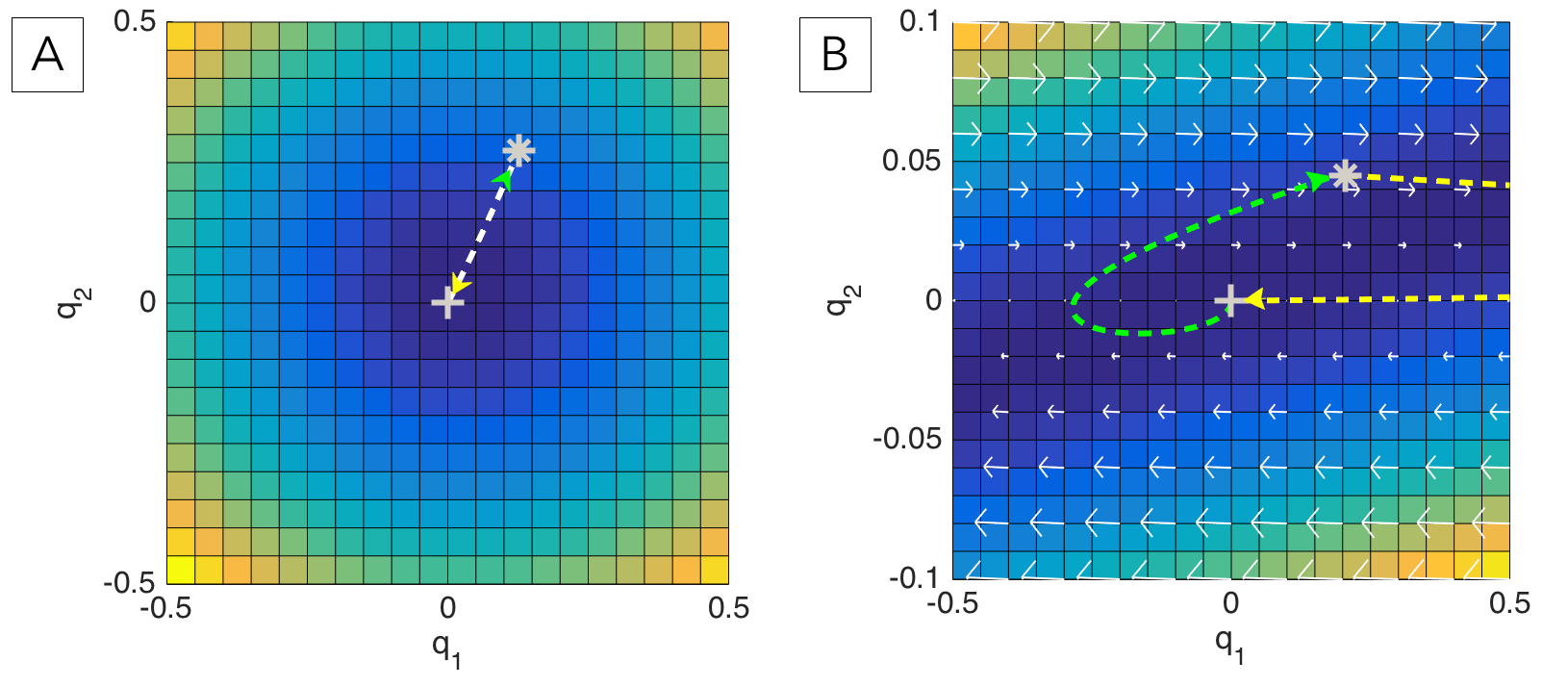}
   \caption{Representation of the scalar potential (color coded) and
     the curl flux force (vector field).  Left panel: For the
     particular (normal) case $w=0$, the residual curl flux force
     vanishes, since the force and the dynamics is fully specified as
     the gradient of the scalar potential. On the other hand (Right
     panel) in the (non-normal) case $w=10$, the system is constrained
     close to the manifold $y=0$ by the potential (the only fixed
     point lies at $(x,y)=(0,0)$ and is weakly stable) and is subject
     to a curl flux determined by the non-conservative part of the
     dynamics ($(W_x,W_y)\simeq(wy,0)$ for
     $\sigma\rightarrow0$). Observe, in particular, that the
     most-probable fluctuating paths to go from the point marked with
     an asterisk to the one marked with a plus sign, is different from
     the one to make the reverse jump.  This reveals the intrinsic
     non-equilibrium (i.e. broken detail balance) nature of the
     problem, while in the case $w=0$ (Left) one trajectory is the
     reverse of the other, i.e. detailed balance is obeyed.
 \label{fig:flux}}
\end{center}
\end{figure}

Once the scalar potential has been determined, we compute the curl
force, $j^\mu/P_{st}$ from Eq.(\ref{eq:F}).  The result is illustrated
in Fig.\ref{fig:flux}: in the $\omega_{ff}=0$ case, the system is
conservative, there is no current nor a curl in the stationary
state. In this case, the most probable path from any given point to
the minimum of the potential is given by the steepest gradient descent
(i.e. the deterministic path) and, vice versa, the most probable path
leading to a fluctuation from the minimum to such a given
point is the time-reversed of the deterministic path (as it is always
the case in equilibrium problems \cite{Graham}).

On the contrary, in the case $\omega_{ff}=10$, the dominant
contribution to the system dynamics is given by the curl flux force
(represented by white arrows in Fig.\ref{fig:pot}).  In this case the
paths clearly deviate from the steepest descent of the scalar
potential and moreover they are no longer reversible, as clearly shown
in Fig.6. Indeed, the figure illustrates that the most likely path to
go from one point to another is quite different from the most likely
path to complete the reverse trip. This reveals the breaking of the
detailed balance condition, illustrating the existence of internal
probability currents, i.e. the non-equilibrium nature of the problem
\cite{WangWaddington,WangPotential}.

%%%%%%%%%%%%%%%%%%%%%%%%%%%%%%%%%

%%%%%%%%%%%%%%%%%%%%%%%%%%%%%%%%%%

\subsection{Stochastic description of the full dynamics}

The mechanism that generates anomalously large fluctuations around the
up-state fixed point has been already rationalized in the previous
sections. However, in spite of this, there is an important aspect of
the actual full dynamics that remains to be clarified. This is the
fact that the full system becomes trapped into a state of extremely
low activity (a down state) for relatively large times (see
Fig.\ref{fig:timeseriesWC}). This phenomenon, which is essential for
the system to exhibit avalanching behavior, is not predicted by the
deterministic dynamics nor by the above weak-noise approach to the
linearized stochastic dynamics. Thus, to shed further light onto it,
one needs to go beyond such approximations, hopefully, constructing an
effective potential for the full stochastic dynamics.

In principle, the effective potential for the full problem might be
calculated using the same method as described above for the linearized
dynamics \cite{Graham,Graham1,Wio,Ja1,Ja2}. However, for the full
problem, owing to the presence of multiplicative/demographic noise, it
is necessary to go beyond the Hamilton-Jacobi approach and consider
the next-to-leading order approximation \cite{Graham1}. We tried to perform such a
calculation, but found a number of mathematical pitfalls; these
technical difficulties make the problem highly non-straightforward and
beyond the scope of the present paper (we leave a detailed
mathematical analysis of this for a future work).

For this reason, we resorted to measure the steady state probability
distribution $P_{st}$ in a purely computational way, and use it to
determine the stationary scalar potential
$V(\Sigma,\Delta)=-\log P_{st}(\Sigma,\Delta)$.  Fig.\ref{fig:2Dpot}
shows that such an effective potential has a deep minimum close to the
origin, as the potentials of stochastic systems with
multiplicative/demographic noise generally do \cite{nature} (see also
\cite{Jenkinson}). Observe that the minimum is not a singularity just
because $h \neq 0$, preventing a true  \new{absorbing state} to exist.
\begin{center}
\textit{}
\begin{figure}[h]
\begin{centering}
\includegraphics[width=0.85\columnwidth]{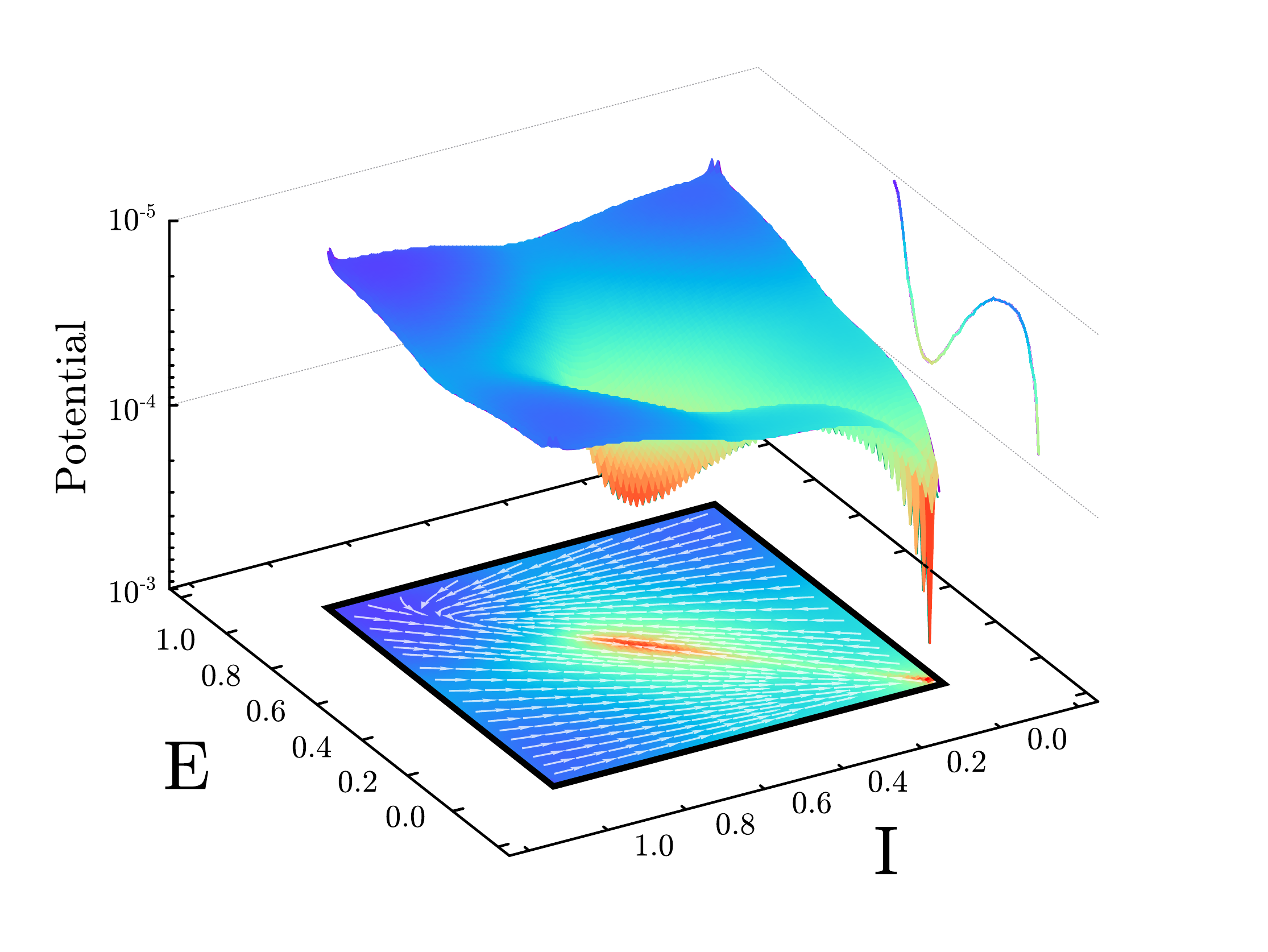}
\par\end{centering}
\centering{}\caption[Stationary scalar potential and curl flux for the
full dynamics.]{Stationary scalar potential of the full dynamics under
  noisy conditions, Eq. (\ref{eq:WCNoise}). Observe the presence of a
  demographic-noise-induced minimum (absent in the previous
  linear-approximation approaches) very close to the
  origin. Interestingly, observe that although the potential tends to
  keep the system dynamical state confined very close to the diagonal
  there exists a strong curl flux force (as shown in the projection on
  the E-I plane) that pulls the system into the
  (multiplicative)noise-induced minimum.  Reddish colors represent the
  different minima. Observe that we have also plotted the projection
  of the potential into the $E$ variable, which is equivalent to the
  potential for the $I$ variable. Parameters: $h=10^{-6}$, $\omega_E=7$,
  $\omega_I=34/5$, $\alpha=0.1$.}
\label{fig:2Dpot}
\end{figure}
\par\end{center}
Moreover, Fig.\ref{fig:2Dpot} shows that, on the one hand, the
potential keeps the dynamics confined very close to the diagonal
($E=I$) and, on the other hand, as soon as the system moves slightly
away from it, the dynamics is governed by the strong curl flux force 
\new{--evaluated numerically form Eq.(\ref{eq:F})--},
which, either pulls the system directly towards the noise-induced
minimum at the origin, or pulls it towards the deterministic
minimum. 

This allows us to rationalize the emerging avalanches: the
system is trapped in the noise-induced minimum; small fluctuations
drive it slightly away from the minimum, and then, the curl flux force
drives the system towards the deterministically stable up
state. However, the system does not dwell long around such an up
state: given that the deterministic dynamics is non-normal and
possibly reactive, and that any tiny stochastic fluctuations drive the
system away from the fixed point while the curl flux force drags it back to
the down state, closing the cycle of an individual avalanche. 

Summing up, the emerging curl flux keeps the system jumping back and
forth (circulating) from the down state to states of larger
activity. This is the mechanism for the emergence of avalanching
behavior in deterministically stable systems, once finite-size
fluctuations are considered.

Finally, let us mention that in Appendix B, we analyze how all this
phenomenology depends on system size (i.e. noise amplitude) and,
moreover, \new{through an effective one-dimensional approach we} explain why the measured avalanche exponents are
compatible with those of a standard random walk.

\subsection{Avalanching without reactivity}

It should be noted that the linear stability analysis of the system at
the deterministic fixed point (in particular, whether the linearized
dynamics is non-normal and reactive, or not) gives a very good idea of
the possibility of having the type of effects discussed in the main
text, e.g. large transient departures from the fixed point in the presence of
fluctuations.

However, it is important to keep in mind that the linearized dynamics
does not include all the information of non-linear systems, and one
could expect some deviations from linear-approximation-based
knowledge, at least in some cases.  

Thus we asked the question whether it could be possible to have
avalanching behavior, as described in the main text, in the case in
which the up-state fixed point is stable, weakly non-normal and non
reactive at all.  Surprisingly, the answer is yes!
 
In order to illustrate this effect, we consider the set of stochastic
Eqs.(\ref{eq:WCNoise}) with a particular choice of the parameter
values (see caption of Fig.\ref{fig:phaseport2}) for which the
linearized dynamics around the fixed point is strongly stable, quasi-normal and
non-reactive but still --very importantly-- the two nullclines are
very close to each other.

As illustrated in Fig.\ref{fig:phaseport2} (left) --even if the fixed
point is strongly stable-- the deterministic field of forces still has
a scar near the diagonal, shear flows exist at both sides of it, and
one can anticipate large excursions away from the up state once
fluctuations are taken into account. As a matter of fact,
Fig.\ref{fig:phaseport2} (right) illustrates that noise amplitudes
above some threshold (larger than in the cases discussed in the main
text) lead the system to exhibit large excursions alternating up and
down states, in a rather intermittent --avalanching like-- way. Thus,
in conclusion, under some circumstances it is possible to have very
nearby nullclines, generating a ``scar'' in the velocity field
--opening the door to avalanching phenomena when noise is included--
even in the absence of non-normal/reactive deterministic dynamics at
the fixed point.  The key ingredient is the proximity of the two
nullclines, regardless of whether when they intersect they do it in
such region of proximity (inducing the two eigenvectors to be almost
degenerate, and the matrix non-normal), or they intersect in a region
whether they are separated, as in Fig.\ref{fig:phaseport2} (Left).  We
propose to call this mechanism, \emph{non-linear reactivity}.

\begin{center}
\begin{figure}[h]
\centering{} \includegraphics[width=0.90\columnwidth]{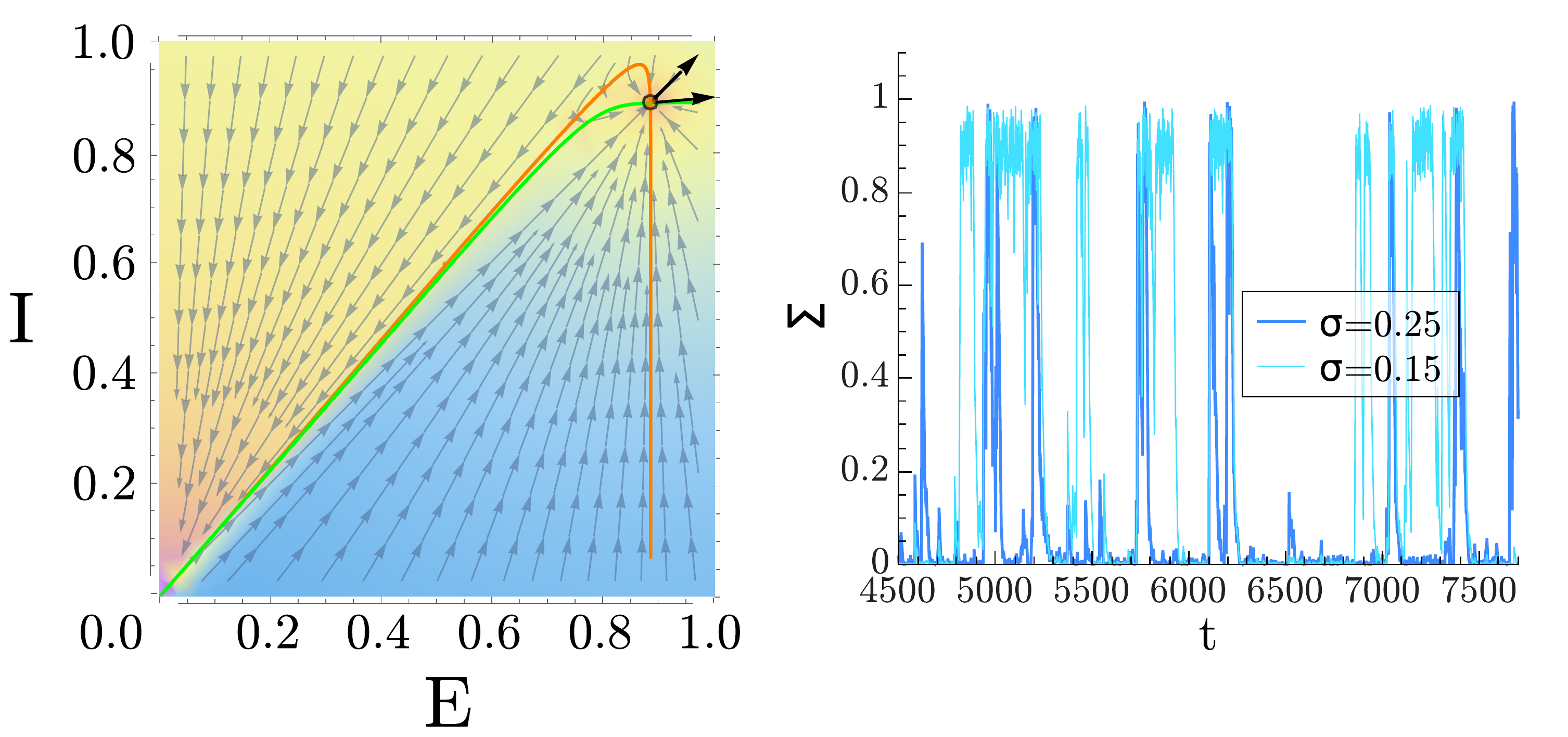}
\caption[Phase portrait of the Wilson-Cowan model]{(Left) E-I phase
  portrait for $\omega_0=3$ --in the active phase-- with
  $\omega_{E}=20$ and $\omega_{I}=17$.  The proximity between the
  nullclines makes it easy that, above certain perturbation, the
  system exhibits large excursions from the up state to the down
  state, even though the non-normality is $\mathcal{NN}=10^{-4}$ and the
  reactivity ${\cal{R}}=-1.08$.  We have called this mechanism,
  \emph{non-linear reactivity}. (b) Time series for two values of the
  noise amplitude $\sigma=0.15$ and $\sigma=0.25$ showing large
  noise-induced fluctuations.  Other parameter values: $\alpha=0.1$
  and $h=10^{-3}$. \label{fig:phaseport2}}
\end{figure}
\par\end{center}

\section{Discussion and conclusions}

The mechanism of amplification of fluctuations was proposed by Murphy
and Miller \cite{MurphyMiller} and later employed by Benayoun et
al. \cite{Benayoun} to explain the origin of avalanching phenomena in
neural systems. It caught our attention as a possible alternative to
criticality to justify the emergence of scale-invariance in
neuronal networks, as first measured empirically by Beggs and Plenz \cite{BP2003}. 

In particular, Benayoun et al. observed that the model of Wilson-Cowan
--very well-known and accepted as a mesoscopic description of neural
dynamics-- including excitatory and inhibitory neuron populations, is
able to describe heavy-tailed distribution of bursts of spontaneous
neuronal activity if it is tuned to a condition of an approximate
balance between excitatory and inhibitory couplings. Importantly, this
occurs without the need for a perfect fine tuning to the very critical
point, at which excitation and inhibition exactly compensate to each
other.

Such a mechanism of ``balanced amplification'' stems from the
existence of a rather peculiar type of stable fixed point at the
deterministic level (i.e. for infinitely large neural populations). In
particular, as we have profusely illustrated here, if the Jacobian
characterizing the linearized deterministic dynamics is a non-normal
matrix --which happens if the conditions for balanced amplification
hold-- then, the corresponding eigenvectors do not form an orthogonal
basis and, as a consequence, some directions can be poorly
represented, implying that the components of vectors in such
directions can become very large in the eigenvectors basis. Moreover,
as discussed along the paper and more specifically in Appendix A,
\new{this implies that the system can be reactive}, meaning that even
if the fixed point is stable, perturbations from it can grow
significantly in modulus before decaying back. We have shown that
reactivity is a more stringent condition than non-normality: all
reactive matrices are non-normal, but the opposite is not true.

\new{This type of anomalous deterministic dynamics induces} the
possibility of transiently evolving away from the deterministic fixed
point.  When finite populations are considered, intrinsic or
demographic noise needs to be taken into account. Such a noise induces
the emergence of a new attractor of the dynamics for very low
activities. Given that such an attractor --a down state-- is
deterministically unstable, fluctuations, \new{can make the system
  escape from it, experiencing large excursions before falling back
  again into the down state, thus giving rise to avalanches of
  activity.}

At the light of all this, one might be tempted to conclude that
non-normality (and, possibly, reactivity) --together with intrinsic
stochasticity-- are the key ingredients for the described phenomena to
emerge.  However, as we have shown, such properties of the linearized
dynamics are not actually necessary conditions. The main required
ingredient in the deterministic dynamics is that the two nullclines
need to be very close in a broad region of the phase portrait --even
if they do not intersect in such a region-- and this is a non-linear
feature.  Actually, we propose to call this \emph{non-linear
  reactivity}.  Still, the easiest way to have very close nullclines
is when the two nullclines intersect when they are almost parallel,
and in such a case the associated linear stability matrix is most
likely non-normal, and possibly reactive.

Thus, the ``balanced amplification'' condition implies that the system
is able to generate some type of ``scar'' or shear flow in the
deterministic velocity field, allowing for very large excursions away
from the up state when the system is perturbed. This together with
demographic noise --which has two effects: (i) taking the system away
from the deterministic fixed point, and (ii) creating a second
noise-induced effective attractor of very low activity-- explains the emergence of
avalanches and large excursions between the down and up states.

In other to further rationalize all this, we have constructed effective
potentials to describe the dynamics. We found, first that, in the case
of balance amplification, the dynamics cannot be simply derived
from the gradient of a scalar potential. There is a very important
contribution from a curl flux (or, equivalently, a vector potential),
that explains why the system circulates back and forth between states
of low an high activity. Importantly, the global potential is a
non-equilibrium one, and the most likely trajectory to jump from one
state to another does not coincide with the most probable one to do the
reverse jump.

Moreover, we have constructed an effective one-dimensional description
of the full problem, that allowed us to understand that the avalanches
around the down state, especially the small ones, can be well
described by the excursions of a random walk around an almost flat
potential. The resulting avalanche exponent values coincide with those
of the unbiased random walks ($\alpha=3/2$ and $\tau=4/3$) and, thus,
are not compatible with the empirically measured ones (in neuronal
tissues) which are better fitted by the exponents of an unbiased
branching process ($\alpha=2$ and $\tau=3/2$). In any case, these are
not perfect power-laws, i.e. the system is not truly scale invariant,
as above some level of activity, the system can get trapped for some
time around the up state, breaking the freely-moving random walk
picture.

On the other hand, preliminary results to be fully developed elsewhere
show that the scenario described here --combining non-normal dynamics
with stochasticity-- can appear in other neural systems (e.g.
including synaptic plasticity rather than inhibition as a chief
regulatory mechanism), clearly illustrating its generality.  Also, the
present mechanism is very likely to play a role in population dynamics
in ecology, a context in which reactivity has been profusely discussed
at a deterministic level (see \cite{Caswell}).

Finally, we hope that this work fosters further analyses of
non-standard deterministic dynamics --in the sense of showing either
non-normal forms of the stability matrix and/or shear flows-- in noisy
dynamical systems of relevance in other fields and in the presence of
more complex networked architectures.

\section{Acknowledgments}
We are grateful to the Spanish-MINECO for financial support (under
grants FIS2013-43201-P and FIS2017-84256-P; FEDER funds). MAM also
acknowledges the support from TeachinParma and the Cariparma
foundation. Last but not least we are happy to acknowledge P. Moretti,
P.  Garrido, S. Suweis, A. Maritan, and especially V. Buendia for
extremely useful comments.

\appendix

\section[\hspace{2cm}Reactivity in non-normal systems]{Reactivity in non-normal systems}

\setcounter{footnote}{0}

\begin{figure}[h]
\begin{center}
\includegraphics[width=0.7\columnwidth]{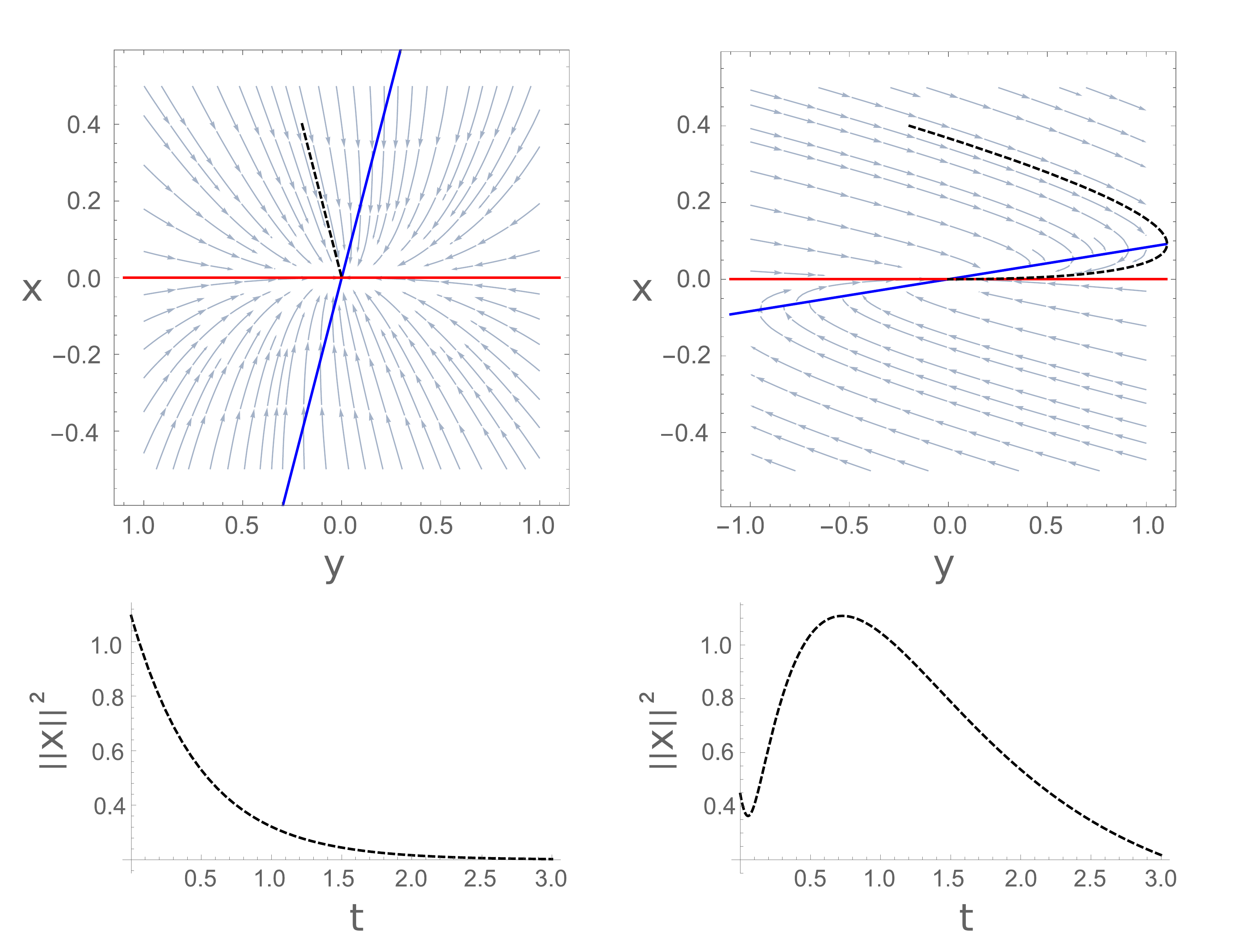}
   \caption[Phase plane]{Phase plane analyses and temporal evolution of the norm of
     a perturbation $\mathbf{x}(t)$, for the linear dynamics
     $\mathbf{J}_1=\left(\begin{array}{cc}-1 & 0.5 \\0 & -1 \end{array}\right)$ (left) and $\mathbf{J}_2=\left( \begin{array}{cc}-1 & 12\\0 & -1 \end{array}\right)$.
     The eigenvalues of the two matrices are the same, $\lambda_{max}^{\mathbf{A}_1}=\lambda_{max}^{\mathbf{J}_2}=-1$,
     but matrix $\mathbf{J}_1$ is not reactive (${\cal{R}}=\lambda_{max}^{H(\mathbf{J}_1)}=\simeq-0.94$), while matrix $\mathbf{J}_2$ is reactive (${\cal{R}}=\lambda_{\max}^{H(\mathbf{J}_2)}\simeq4.52$). Note
     that both $\mathbf{J}_1$ and $\mathbf{J}_2$ are non-normal; normal matrices are always non reactive. \label{fig:Casw}}
\end{center}
\end{figure}

The stability of the fixed point $P$ of a given dynamical system is
determined evaluating the eigenvalues of its associated Jacobian
matrix $\mathbf{J}$ corresponding to the linearization of the dynamics
close to the fixed point \cite{Ott}.  This gives information about the
system's \emph{long-term} asymptotic behavior: if all the eigenvalues are real and
negative, then, in the large-time limit, the perturbed system will
return to the fixed point $P$ with certainty. However, this does not
give information about the \emph{short-term} dynamics of the
linearized dynamics, which --in some cases-- can be affected by long
transients. In Fig.\ref{fig:Casw} we illustrate a (non-normal) stable
linear system can transiently evolve away from its stable fixed point
or, in other words, a perturbation can transiently grow before
decaying to zero \cite{Caswell}. Actually, the temporal evolution of a
perturbation $\Vert\mathbf{x}(t)\Vert$ is described by:
\begin{equation}
  \frac{d\Vert\mathbf{x}\Vert}{dt}=\frac{d\sqrt{\mathbf{x}^T\mathbf{x}}}{dt}=\frac{\mathbf{x}^T(d\mathbf{x}/dt)+(d\mathbf{x}/dt)^T\mathbf{x}}{2\Vert\mathbf{x}\Vert}=\frac{\mathbf{x}^T H(\mathbf{J})\mathbf{x}}{\Vert\mathbf{x}\Vert}
\end{equation}
where $H(\mathbf{J})\equiv(\mathbf{J}+\mathbf{J}^T)/2$ is called the
\emph{Hermitian part} (or \emph{symmetric part}) of $\mathbf{J}$.  The
maximum amplification rate immediately following the perturbation,
specifying the \emph{reactivity} (or \emph{Rayleigh quotient} of $H$)
 of the system can thus be written as:
\begin{equation}
 {\cal{R}}= \lambda_{max}^{H}= \max_{\Vert\mathbf{x}_0\Vert\neq 0} \left(\frac{1}{\Vert\mathbf{x}\Vert}\frac{d\Vert\mathbf{x}\Vert}{dt}\right)\bigg|_{t=0} =  \max_{\Vert\mathbf{x}_0\Vert\neq0} \frac{\mathbf{x}_0^T H(\mathbf{J})\mathbf{x}_0}{\Vert\mathbf{x}_0\Vert},
\end{equation}
which is nothing but the maximum eigenvalue of $H(\mathbf{J})$
\cite{matrix}.  This means that the strongest magnification that can
occur must take place in the eigenspace belonging to the eigenvalue
with the largest modulus --for all other vectors the magnification
will be smaller-- and the magnification is equal to the maximum
eigenvalue of the Hermitian part itself.  Thus, in summary, while the
asymptotic behavior is determined by the maximum eigenvalue of matrix
$\mathbf{J}$, the transient behavior is determined by the maximum eigenvalue of
the Hermitian part $H(\mathbf{J})$.

If $\mathbf{J}$ is not symmetric, then $\mathbf{J}\neq H(\mathbf{J})$ and, thus, it is
possible that the maximum eigenvalue of $\mathbf{J}$ is negative, while the
maximum eigenvalue of $H(\mathbf{J})$, $\lambda_{max}^{H}$, is
positive, meaning that the equilibrium specified by $\mathbf{J}$ is
\emph{stable but reactive} \cite{Caswell}.  

Observe, in particular, that Rayleigh's principle does not hold for
non-normal matrices\footnote{Note that $H(\mathbf{J})$ is Hermitian
  and all Hermitian matrices are normal.} , i.e. if $\mathbf{J}$ is
non-normal, then it is not true that the strongest amplification must
occur in the direction of the eigenvector corresponding to the largest
eigenvalue, thus, there can exist strongly-amplified directions.
In particular, the Schur theorem \cite{Horn,Trefethen} states that non-normal
matrices can always be transformed in upper triangular matrices,
having the eigenvalues on the diagonal and a ``feedforward'' dynamics
outside the diagonal, which cannot be neglected and whose relative
weight determines the strength of the non-normality.

Finally, let us remark that reactivity is a more stringent condition
than non-normality: reactive matrices are always non-normal, but the
opposite is not true.  Let us clarify the interplay between reactivity
and non-normality through a simple example. Take the matrix
$\mathcal{J}=\left(\begin{array}{cc} -1 & b\\0 & -d\end{array}\right)$
and its Hermitian
$\mathcal{H}=\left(\begin{array}{cc}-1 & b/2\\b/2 &
    -d\end{array}\right)$ whose maximum eigenvalue is
$\lambda^H_{max}=-1-d+\sqrt{1+b^2-2d+d^2}$. The condition $b^2>4d$ specifies reactivity
whereas $b>0$ is the condition for non-normality. Thus reactivity is a more stringent
condition than non-normality.
%For instance, Fig.4d shows a non-normal fixed point without any sign of reactivity.

\section[\hspace{2cm}Effective one-variable description: stochastic tunneling and
  the origin of the avalanche exponents]{Effective one-variable description: stochastic tunneling and
  the origin of the avalanche exponents}

Despite the fact that the main ingredients of the avalanching
phenomenon have been carefully scrutinized in previous sections, we
still do not have a transparent explanation for the computational
observed avalanche exponent values (compatible with the statistics of
standard unbiased random walks), nor a detailed analysis on how the
phenomenon depends on noise amplitude (i.e. on system size).

Aimed at shedding further light on these issues, in this appendix we
present a one-dimensional effective description to the full
two-dimensional dynamics Eq.(\ref{eq:WCNoise}).

\begin{center}
\begin{figure}[h]
\centering{} \includegraphics[width=0.98\columnwidth]{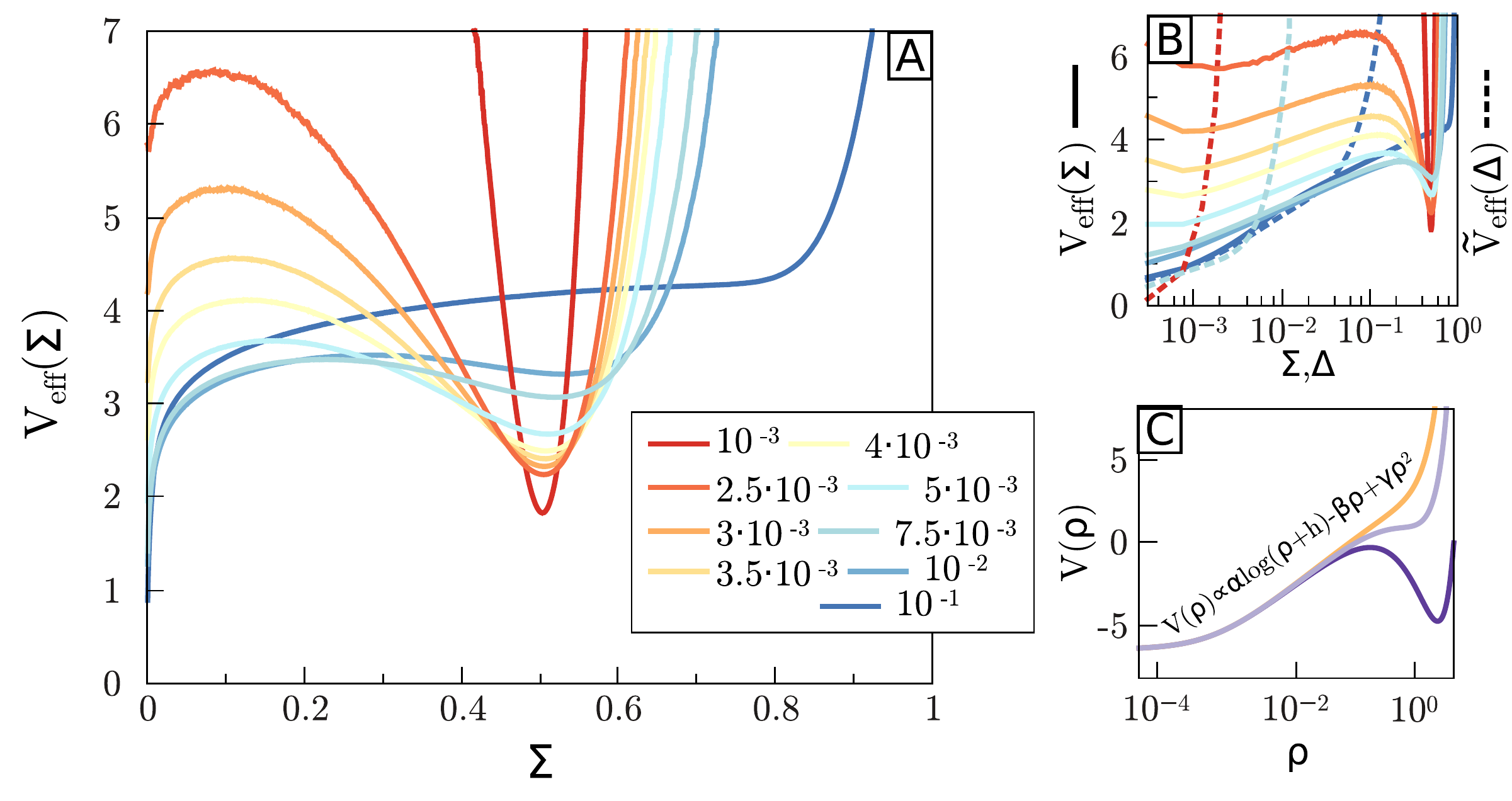}
\caption[Histogram of the $\Sigma$-signal of one Wilson-Cowan column
for different values of the noise amplitude]{(a) Effective potential
  for the overall activity $\Sigma$ for different values of the noise
  amplitude ($\sigma$, see legend). Its shape changes from a single
  potential well for low noise (dark red curve) to a bistable
  situation alternating between up and down states and, finally, to a
  new single potential peaked around the down state for larger noise
  (blue curve). (b) Same potential but plotted in double logarithmic
  scale.  Dashed lines correspond to the potential
  $\tilde V_{eff}(\Delta)$, showing parabolic-like potentials centered
  in $\Delta=0$. (c) Analytically computed potentials derived from the
  one-dimensional effective equation for $\rho$, strongly resembling
  $V_{eff}(\Sigma)$. Parameter values: $\omega_{E}=7$,
  $\omega_{I}=\frac{34}{5},\,\alpha=0.1,\,h=10^{-6}$. \label{fig:potentials}
}
\end{figure}
\end{center}

We have already computed the bivariate probability distribution
$P(\Sigma,\Delta)=\exp(-V(\Sigma,\Delta))$.  Marginalizing over
$\Delta$, one obtains
$P(\Sigma)=\int_{0}^{\infty}P(\Sigma,\Delta)d\Delta$, from which one
can define an effective one-dimensional potential,
$V_{eff}(\Sigma)=-\ln P(\Sigma)$, as illustrated in
Fig.\ref{fig:potentials}A for different values of the noise amplitude.
Observe, in particular, that the effective potential becomes more and
more peaked around the deterministic solution ($\Sigma\simeq0.5$) as the
noise amplitude is reduced (i.e. for large systems sizes), while for bigger noise amplitude a new noise-induced minimum emerges near the origin.
This explains the unusually large permanence times of the original
two-variable system into low activity regimes, which is ultimately
responsible for the avalanching behavior.

Moreover, in Fig.\ref{fig:potentials}B --where a double logarithmic
scale is employed-- the structure of the potential $V_{eff}(\Sigma)$
can be further inspected, revealing the presence of a logarithmic
decay.  Also, in Fig.\ref{fig:potentials}B we plot --with dashed
lines-- the complementary potential $\tilde{V}_{eff}(\Delta)$ obtained
from $P(\Delta)= \int_{0}^{\infty}P(\Sigma,\Delta)d\Sigma$, through
$\tilde{V}_{eff}(\Delta)=-\log(P(\Delta))$.  Observe that for all
noise intensities the potential has a minimum around $\Delta=0$,
indicating that the dynamics spends most of the time close to the
diagonal $E=I$.

To have some additional understanding of the origin of the
noise-induced minimum of $V_{eff}(\Sigma)$ near the origin one can
  imagine (just for argument's sake) that the dynamics of the system
  was strictly constrained to evolve along the $E=I$ diagonal, and let
  us define a density variable $\rho=E=I$), which implies fixing
  $\Delta=0$.   Performing a Taylor expansion and keeping only leading
  terms in Eq.(\ref{eq:WCSigma}) gives rise to
\begin{equation}
\dot{\rho}=h+a\rho-b\rho^{2}+\tilde{\sigma}\sqrt{(\rho+h')}\eta,\label{eq:cp}
\end{equation}
with $a=\left(-\alpha-h+\omega_{0}\right)$, $b=\omega_{0}$, 
$\tilde{\sigma}=\sigma\sqrt{(\omega_{0}+\alpha-h)}$, 
$h'=h/(\omega_{0}+\alpha-h)$, and $\eta$ is, as above, a Gaussian white variable.

The effective potential resulting from this simple one-dimensional
dynamics can be easily calculated (by writing the associated
Fokker-Planck equation and computing its steady state)
\begin{equation}
 V(\rho)=\log(\rho+h')-\frac{2}{\tilde\sigma^2}\int\frac{a\rho+b\rho^2+h}{\rho+h'}.
 \label{eq:1Dpot}
\end{equation}
A sketch of this potential is shown in Fig.\ref{fig:potentials}C.
Observe that it reproduces all the key features of the computationally
determined one in Fig.\ref{fig:potentials}B.

Finally, let us remark that this approximation is not exact: actually,
it is not true in the actual dynamics that the system remains confined
to the diagonal (i.e. $E \neq I$, and thus $\Delta \neq 0$); indeed,
one of the key ingredients of the complete dynamics is the shear flow
occurring out of the diagonal, i.e. the non-conservative
contribution from the curl flux.  Thus, in order to mimic the effect
of off-diagonal shear flow-- we additionally introduce a ``stochastic
tunneling effect'' in our one-variable ($\rho$) effective description,
such that the system can spontaneously jump from the down state to the
up one and the other way around with some fixed probabilities ($p$).
In this way, we have constructed a new``effective'' one-dimensional
dynamics for the full (two-dimensional) problem. While the resulting
potential keeps essentially the same shape as shown in
Fig.\ref{fig:potentials}B, we have computed the avalanche
distribution using this effective dynamics; results are illustrated
in Fig.\ref{fig:quasi-dimensional}. In particular, the distributions
are consistent --at least for small avalanches-- with the first-return
statistics of unbiased random walkers, justifying the computationally
measured exponents in the main text. For larger avalanches, there is
always a bump in the probability distributions --much as in the neural
model-- corresponding to avalanches remaining for a long time near the
up fixed point.

\begin{center}
\textit{}
\begin{figure}[h]
\begin{centering}
\includegraphics[width=0.65\columnwidth]{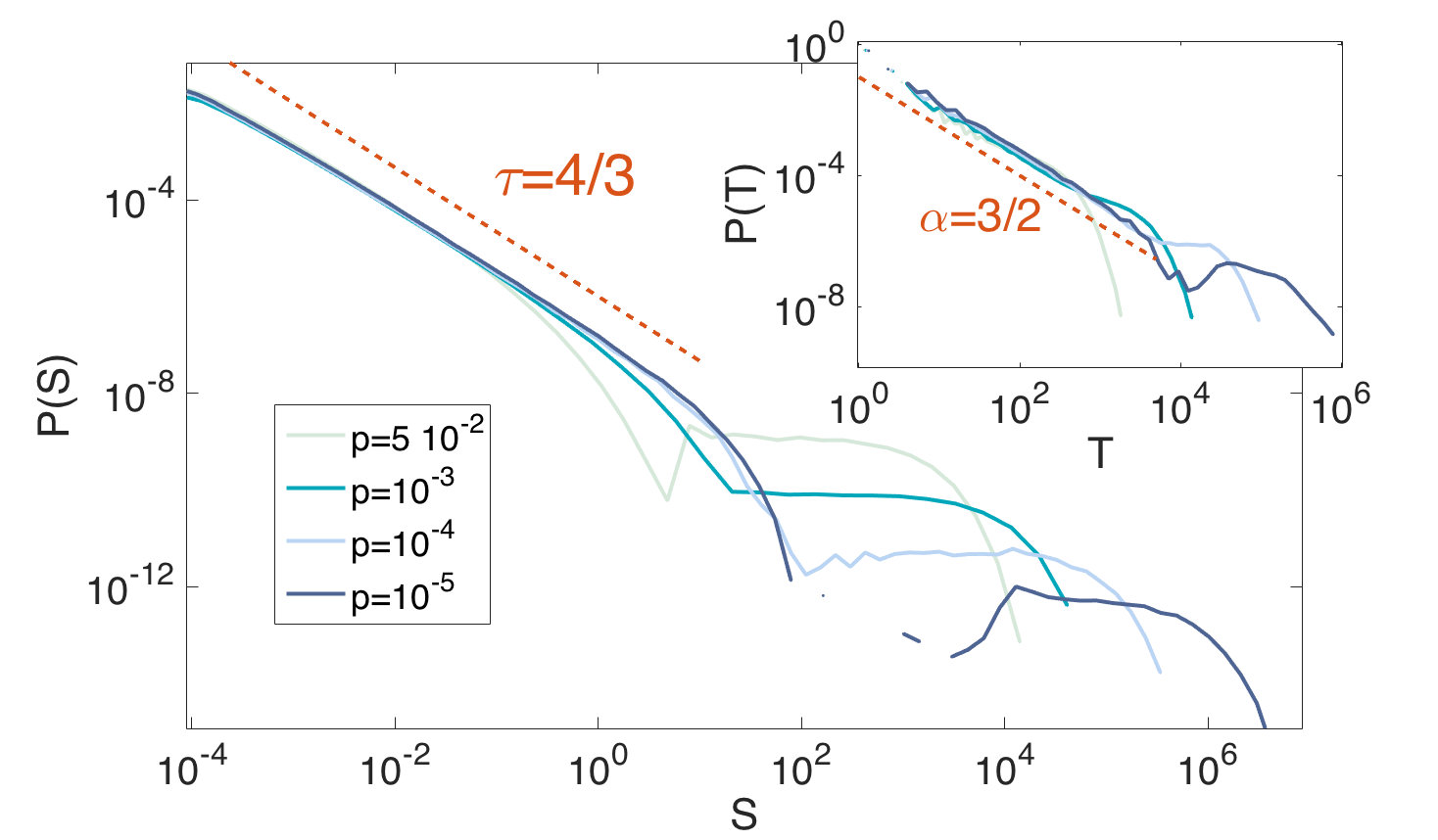}
\par\end{centering}
\centering{}\caption[Avalanches for the  effective one-dimensional
model.]{ Avalanche size distribution $P(S)$ (main) and duration
  distribution $P(T)$ (inset) for different values of $p$
in the effective one-dimensional model. 
 Observe that the exponent values are compatible with those of and
 unbiased random walk for small avalanche sizes. 
For larger sizes, deviations from perfect scaling occur (as they do in
the original two-dimensional system), in particular ``bumps''
apper. These reveal the existence of a
more complicated dynamics with jumps to the up state, where the
system can get trapped for some time, before the avalanche ends.
Parameter values: $a=0.5$, $b=1$, $\sigma=10^{-3}$, $h=10^{-3}$, $T=2.5\cdot10^{-4}$.
\label{fig:quasi-dimensional}}
\end{figure}
\par\end{center}

\newpage

\section*{References}
\bibliographystyle{iopart-num}
\bibliography{BALANCED}

\providecommand{\newblock}{}
\begin{thebibliography}{10}
\expandafter\ifx\csname url\endcsname\relax
  \def\url#1{{\tt #1}}\fi
\expandafter\ifx\csname urlprefix\endcsname\relax\def\urlprefix{URL }\fi
\providecommand{\eprint}[2][]{\url{#2}}
% Bibliography created with iopart-num v2.1
% /biblio/bibtex/contrib/iopart-num

\bibitem{induced}
Horsthemke W 1984 Noise induced transitions {\em Non-Equilibrium Dynamics in
  Chemical Systems\/} (Berlin, Heidelberg: Springer) pp 150--160

\bibitem{Ojalvo-review}
Lindner B, Garc{\i}a-Ojalvo J, Neiman A and Schimansky-Geier L 2004 {\em Phys.
  Rep.\/} {\bf 392} 321--424

\bibitem{Sagues}
Sagu{\'e}s F, Sancho J~M and Garc{\'\i}a-Ojalvo J 2007 {\em Rev. Mod. Phys.\/}
  {\bf 79} 829

\bibitem{McKane}
McKane A~J and Newman T~J 2005 {\em Phys. Rev. Lett.\/} {\bf 94} 218102

\bibitem{Wallace}
Wallace E, Benayoun M, Van~Drongelen W and Cowan J~D 2011 {\em PLoS One\/} {\bf
  6} e14804

\bibitem{Hidalgo}
Hidalgo J, Seoane L~F, Cort{\'e}s J~M and Mu{\~n}oz M~A 2012 {\em PLoS One\/}
  {\bf 7} e40710

\bibitem{Biancalani}
Biancalani T, Dyson L and McKane A~J 2014 {\em Phys. Rev. Lett.\/} {\bf 112}
  038101

\bibitem{Genovese}
Genovese W, Mu{\~n}oz M~A and Sancho J~M 1998 {\em Phys. Rev. E\/} {\bf 57}
  R2495

\bibitem{restless}
Raichle M~E 2011 {\em Brain Connect.\/} {\bf 1} 3--12

\bibitem{Fox}
Fox M~D and Raichle M~E 2007 {\em Nat. Rev. Neurosci.\/} {\bf 8} 700

\bibitem{Arieli}
Arieli A, Sterkin A, Grinvald A and Aertsen A 1996 {\em Science\/} {\bf 273}
  1868--1871

\bibitem{Deco}
Deco G, Jirsa V~K and McIntosh A~R 2011 {\em Nat. Rev. Neurosci.\/} {\bf 12} 43

\bibitem{Schuster}
Plenz D and Niebur E 2014 {\em {Criticality in neural systems}\/} (New Jersey:
  John Wiley \& Sons)

\bibitem{BP2003}
Beggs J~M and Plenz D 2003 {\em J. Neurosci.\/} {\bf 23} 11167--11177

\bibitem{Plenz2007}
Plenz D and Thiagarajan T~C 2007 {\em Trends Neurosci.\/} {\bf 30} 101--110

\bibitem{Beggs2008}
Beggs J~M 2008 {\em Philos. Trans. A Math. Phys. Eng. Sci.\/} {\bf 366}
  329--343

\bibitem{Beggs2012}
Friedman N, Ito S, Brinkman B~A, Shimono M, DeVille R~L, Dahmen K~A, Beggs J~M
  and Butler T~C 2012 {\em Phys. Rev. Lett.\/} {\bf 108} 208102

\bibitem{Arcangelis2012}
Lombardi F, Herrmann H, Perrone-Capano C, Plenz D and {De Arcangelis} L 2012
  {\em Phys. Rev. Lett.\/} {\bf 108} 228703

\bibitem{Petermann2009}
Petermann T, Thiagarajan T~C, Lebedev M~A, Nicolelis M~A, Chialvo D~R and Plenz
  D 2009 {\em Proc. Natl. Acad. Sci. USA\/} {\bf 106} 15921--15926

\bibitem{Hahn2010}
Hahn G, Petermann T, Havenith M~N, Yu S, Singer W, Plenz D and Nikoli{\'c} D
  2010 {\em J. Neurophysiol.\/} {\bf 104} 3312--3322

\bibitem{Palva2012}
Palva J~M, Zhigalov A, Hirvonen J, Korhonen O, Linkenkaer-Hansen K and Palva S
  2013 {\em Proc. Natl. Acad. Sci. USA\/} {\bf 110} 3585--3590

\bibitem{Plenz2015-pyramidal}
Bellay T, Klaus A, Seshadri S and Plenz D 2015 {\em Elife\/} {\bf 4} e07224

\bibitem{MAM-review}
Mu{\~n}oz M~A 2017 {\em arXiv preprint arXiv:1712.04499. To appear in Rev. Mod.
  Phys.\/}

\bibitem{Chialvo2010}
Chialvo D~R 2010 {\em Nat. Phys.\/} {\bf 6} 744--750

\bibitem{Mora-Bialek}
Mora T and Bialek W 2011 {\em J. Stat. Phys.\/} {\bf 144} 268--302

\bibitem{Millman2010}
Millman D, Mihalas S, Kirkwood A and Niebur E 2010 {\em Nat. Phys.\/} {\bf 6}
  801--805

\bibitem{Levina2009}
Levina A, Herrmann J~M and Geisel T 2009 {\em Phys. Rev. Lett.\/} {\bf 102}(11)
  118110

\bibitem{Levina2007}
Levina A, Herrmann J~M and Geisel T 2007 {\em Nat. Phys.\/} {\bf 3} 857--860

\bibitem{Bonachela2010}
Bonachela J~A, {De Franciscis} S, Torres J~J and Mu{\~n}oz M~A 2010 {\em J.
  Stat. Mech. Theory Exp.\/} {\bf 2010} P02015

\bibitem{Rubinov2011}
Rubinov M, Sporns O, Thivierge J~P and Breakspear M 2011 {\em PLoS Comput.
  Biol.\/} {\bf 7} e1002038

\bibitem{LG-PNAS}
di~Santo S, Villegas P, Burioni R and Mu{\~n}oz M~A 2018 {\em Proc. Natl. Acad.
  Sci. USA\/}  201712989

\bibitem{Touboul2010}
Touboul J and Destexhe A 2010 {\em PLoS One\/} {\bf 5} e8982

\bibitem{Touboul2}
Touboul J and Destexhe A 2017 {\em Phys. Rev. E\/} {\bf 95} 012413

\bibitem{SOB}
di~Santo S, Burioni R, Vezzani A and Mu{\~n}oz M~A 2016 {\em Phys. Rev.
  Lett.\/} {\bf 116} 240601

\bibitem{neutral-neural}
Martinello M, Hidalgo J, Maritan A, di~Santo S, Plenz D and Mu{\~n}oz M~A 2017
  {\em Phys. Rev. X\/} {\bf 7} 041071

\bibitem{Izhikevich}
Izhikevich E~M 2007 {\em Dynamical systems in neuroscience\/} (Cambridge, MA:
  MIT press)

\bibitem{vVS}
van Vreeswijk C and Sompolinsky H 1996 {\em Science\/} {\bf 274} 1724--1726

\bibitem{Brunel2000}
Brunel N 2000 {\em J. Comput. Neurosci.\/} {\bf 8} 183--208

\bibitem{Lim}
Lim S and Goldman M~S 2013 {\em Nat. Neurosci.\/} {\bf 16} 1306

\bibitem{Benayoun}
Benayoun M, Cowan J~D, van Drongelen W and Wallace E 2010 {\em PLoS Comput.
  Biol.\/} {\bf 6} e1000846

\bibitem{MurphyMiller}
Murphy B~K and Miller K~D 2009 {\em Neuron\/} {\bf 61} 635--648

\bibitem{Wilson1972}
Wilson H~R and Cowan J~D 1972 {\em Biophys. J.\/} {\bf 12} 1

\bibitem{BorisyukKirillov}
Borisyuk R~M and Kirillov A~B 1992 {\em Biol. Cyber.\/} {\bf 66} 319--325

\bibitem{HoppensteadtIzhikevich}
Hoppensteadt F~C and Izhikevich E~M 2012 {\em Weakly connected neural
  networks\/} vol 126 (New York: Springer)

\bibitem{36WC}
Destexhe A and Sejnowski T~J 2009 {\em Biol. Cyber.\/} {\bf 101} 1--2

\bibitem{WC-review}
Cowan J~D, Neuman J and van Drongelen W 2016 {\em J. Math. Neurosci.\/} {\bf 6}
  1

\bibitem{Fanelli}
Zankoc C, Biancalani T, Fanelli D and Livi R 2017 {\em Chaos Solitons
  Fractals\/} {\bf 103} 504--512

\bibitem{Gardiner}
Gardiner C~W 2004 {\em {Handbook of stochastic methods: for physics, chemistry
  and the natural sciences; 3rd ed.}\/} Springer Series in Synergetics (Berlin:
  Springer)

\bibitem{vanKampen}
Van~Kampen N~G 1992 {\em Stochastic processes in physics and chemistry\/}
  (North-Holland: Elsevier)

\bibitem{logarithmicpot}
di~Santo S, Villegas P, Burioni R and Mu{\~n}oz M~A 2017 {\em Phys. Rev. E\/}
  {\bf 95} 032115

\bibitem{avalanches}
Mu{\~n}oz M~A, Dickman R, Vespignani A and Zapperi S 1999 {\em Phys. Rev. E\/}
  {\bf 59} 6175

\bibitem{Horn}
Horn R~A and Johnson C~R 1990 {\em Matrix analysis\/} (Cambridge: Cambridge
  Univ. Press)

\bibitem{Trefethen}
Trefethen L~N and Embree M 2005 {\em Spectra and pseudospectra: the behavior of
  nonnormal matrices and operators\/} (Princeton: Princeton Univ. Press)

\bibitem{Hydrodynamics}
Farrell B~F and Ioannou P~J 1993 {\em ‎J. Atmospheric Sci.\/} {\bf 50}
  4044--4057

\bibitem{Magnetohydrodynamics}
Borba D, Riedel K, Kerner W, Huysmans G, Ottaviani M and Schmid P 1994 {\em
  Phys. Plasmas\/} {\bf 1} 3151--3160

\bibitem{pipe}
Kerswell R 2005 {\em Nonlinearity\/} {\bf 18} R17

\bibitem{Trefe2}
Trefethen L~N, Trefethen A~E, Reddy S~C and Driscoll T~A 1993 {\em Science\/}
  {\bf 261} 578--584

\bibitem{Uncertain}
Hinrichsen D and Pritchard A 1994 {\em Math. Res.\/} {\bf 77} 159--159

\bibitem{Caswell}
Verdy A and Caswell H 2008 {\em Bull. Math. Biol.\/} {\bf 70} 1634--1659

\bibitem{Lasers}
Siegman A 1995 {\em Appl. Phys. B\/} {\bf 60} 247--257

\bibitem{Q}
Bender C~M 2007 {\em Rep. Prog. Phys.\/} {\bf 70} 947

\bibitem{non-HermitianQM}
Hatano N and Nelson D~R 1996 {\em Phys. Rev. Lett.\/} {\bf 77} 570

\bibitem{Risken}
Risken H 1996 Fokker-planck equation {\em The Fokker-Planck Equation\/}
  (Springer) pp 63--95

\bibitem{Dotsenko}
Dotsenko V, Macio{\l}ek A, Vasilyev O and Oshanin G 2013 {\em Phys. Rev. E\/}
  {\bf 87} 062130

\bibitem{Wu}
Wu W and Wang J 2014 {\em J. Chem. Phys.\/} {\bf 141} 105104

\bibitem{Puglisi}
Sarracino A, Villamaina D, Gradenigo G and Puglisi A 2010 {\em EPL (Europhysics
  Letters)\/} {\bf 92} 34001

\bibitem{Graham1}
Graham R and T{\'e}l T 1984 {\em Phys. Rev. Lett.\/} {\bf 52} 9

\bibitem{Graham}
Graham R and T{\'e}l T 1984 {\em J. Stat. Phys.\/} {\bf 35} 729--748

\bibitem{Wio}
Wio H~S 2013 {\em Path integrals for stochastic processes: An introduction\/}
  (New York: World Scientific)

\bibitem{Ja1}
Jauslin H 1987 {\em Physica A\/} {\bf 144} 179--191

\bibitem{Ja2}
Jauslin H 1986 {\em J. Stat. Phys.\/} {\bf 42} 573--585

\bibitem{Arfken}
Arfken G~B, Weber H~J and Harris F~E 2013 Mathematical methods for physicists

\bibitem{WangWaddington}
Wang J, Zhang K, Xu L and Wang E 2011 {\em Proc. Natl. Acad. Sci. USA\/} {\bf
  108} 8257--8262

\bibitem{WangPotential}
Wang J, Xu L and Wang E 2008 {\em Proc. Natl. Acad. Sci. USA\/} {\bf 105}
  12271--12276

\bibitem{nature}
Mu{\~n}oz M~A 1998 {\em Phys. Rev. E\/} {\bf 57} 1377

\bibitem{Jenkinson}
Jenkinson G and Goutsias J 2014 {\em PLoS Comput. Biol.\/} {\bf 10} e1003411

\bibitem{Ott}
Ott E 2008 {\em Chaos in dynamical systems\/} (Cambridge: Cambridge Univ.
  Press)

\bibitem{matrix}
Horn R~A and Johnson C~R 2013 {\em Matrix analysis\/} (Cambridge: Cambridge
  Univ. Press)

\end{thebibliography}

\end{document}